\documentclass[a4paper,11pt,
]{article}
\usepackage{amsmath,amssymb}
\usepackage[dvips]{color}
\usepackage{cite}
\usepackage{hangcaption}
\usepackage{amsmath,amssymb}
\usepackage[dvips]{color}
\usepackage{cite}
\usepackage{eepic}
\usepackage{epic}
\usepackage{hangcaption}
\usepackage{amsmath}
\usepackage{amssymb}
\usepackage{amscd}
\usepackage{amsthm}
\usepackage[dvipdfmx,%
bookmarks=true,%
bookmarksnumbered=true,%
setpagesize=false,%
pdfkeywords={TeX; dvipdfmx; hyperref; color;}]{hyperref}
\usepackage{hyperref}
\def\cases{\left\{\begin{array}{ll}}
\def\endcases{\end{array}\right.}

\def\bigtimes{\mathop{\mbox{\Large $\times$}}}
\begin{document}
\setcounter{page}{1}
%\twocolumn[
\vskip1.5cm
%\centerline
\begin{center}
{\LARGE \bf 
The two envelopes paradox in non-Bayesian and Bayesian statistics
}
%Explanation which is the easiest to understand}
%Is Statistics Based on Kolmogorov's Probability Theory?}
%\\
%{\LARGE \bf
%in the Quantum Mechanical World View}
\vskip0.5cm
%\begin{center}
{\rm
\large
Shiro Ishikawa
%,  Kohshi Kikuchi
}
\\
\vskip0.2cm
\rm
\it
\it
Department of Mathematics, Faculty of Science and Technology,
Keio University,
\\ 
3-14-1, Hiyoshi, Kouhoku-ku Yokohama, 223-8522, Japan.
{E-mail:
ishikawa@math.keio.ac.jp}%
%\newline
%\newline
%Department of Applied Molecular Bioscience, Graduate School of Medicine, Yamaguchi University
%\\
%2-16-1 Tokiwadai, Ube 755-8611, Japan,
%E-mail:
%kohshi.kikuchi@gmail.com
\end{center}
\par
\rm
\vskip0.4cm
\par
\noindent
{\bf Abstract}
\normalsize
%\vskip0.5cm
\par
\noindent
The purpose of this paper is to clarify
the (non-Bayesian and Bayesian ) two-envelope problems in terms of quantum language (or, measurement theory), which was recently proposed as a  linguistic turn of quantum mechanics (with the Copenhagen interpretation). 
The two envelopes paradox is only a kind of high school student's probability puzzle, and it may be exaggerated to say that this is an unsolved problem.
However, since we are convinced that quantum language is just statistics of the future, we believe that there is no clear answer without the description by quantum language. In this sense, the readers are to find that quantum language provides the final answer (i.e., the easiest and deepest understanding ) to the two envelope-problems in both non-Bayesian and Bayesian statistics.
Also, we add the discussion about St. Petersburg two-envelope paradox.

\par
\par
\vskip0.3cm
\par
%\noindent
%{\bf Keywords}: the Copenhagen Interpretation, 
%Operator Algebra,
%Quantum and Classical Measurement Theory,
%Bayesian statistics
%%, Bertrand paradox
%%Kalman Filter.
%%%
%Fisher Maximum Likelihood Method,
%Regression Analysis,
%Philosophy of Statistics
\par
%\vskip0.5cm
\par
%]

\par
%\vskip2.0cm
%\par
%%
%%\newtheorem{Def}{Definition}[chapter]
%\newtheorem{The}[Def]{Theorem}
%\newtheorem{Exa}[Def]{Example}
%\newtheorem{Not}[Def]{Notation}
%\newtheorem{Rem}[Def]{Remark}
%\newtheorem{Prp}[Def]{Proposition}
%\newtheorem{Lem}[Def]{Lemma}
%\newtheorem{Ans}[Def]{Answer}
%\newtheorem{Prb}[Def]{Problem}
%\newtheorem{Cor}[Def]{Corollary}
%\newtheorem{Met}[Def]{Method}
%\newtheorem{Sit}[Def]{Situation}
%\newtheorem{Hyp}[Def]{Hypothesis}
%%}111111111111111111
\def\Cal{\cal}
\def\bigstimes{\text{\large $\: \boxtimes \,$}}

%1111111111111111111
%\vskip1.5cm
\par
\noindent
\section{
Introduction}
%\par
%
%
%\par
%\noindent
\subsection{
Two-envelope paradox
%(
%the two-envelope paradox
%%and the St. Petersburg two-envelope paradox
%)
}
\rm
\par
\noindent
\par
In what follows, we firstly introduce the two-envelope problem
({\it cf.} {\cite{Mart, Nale}}),
which is well known
as a kind of high school students' mathematical puzzle.

\noindent
\par
\noindent
{\bf Problem 1}
\rm
[The two envelope problem].
The host presents you with a choice between two envelopes
(i.e.,
Envelope A
and
Envelope B).  You know one envelope contains twice as much money as the other, but you do not know which contains more.  You choose randomly (by a fair coin toss) one envelope,
for example,
call it Envelope A.
Suppose that you find $\alpha$ dollars inside your envelope $A$.
Now the host says
"You are offered the options of keeping your A or switching to my B". What should you do? 
\par \vskip0.3cm
\par
\noindent
[(P1):Why is it paradoxical?].
You reason that, with probability 1/2, the other envelope $B$ has either $\alpha/2$ or $2\alpha$ dollars. Thus the expected value (denoted $E_{\mbox{\small other}}(\alpha)$ at this moment) of the other envelope is 
\begin{align}
E_{\mbox{\small other}}
(\alpha)=(1/2)(\alpha/2) + (1/2)(2\alpha)=1.25\alpha
\label{eq1}
\end{align}
This is greater than the $\alpha$ in your current envelope $A$. Therefore,
you should switch to B. But this seems clearly wrong, as your information about A and B is symmetrical. 
This is the famous two-envelope paradox
(i.e.,
"The Other Person's Envelope is Always Greener"
).

%\noindent
%\par
%\noindent
%{\bf Problem 1}
%\rm
%[The two envelope problem].
%You are presented with a choice between two envelopes
%(i.e.,
%Envelope A
%and
%Envelope B).  You know one envelope contains twice as much money as the other, but you do not know which contains more.  You choose randomly (by a fair coin toss) one envelope,
%for example,
%call it Envelope A.
%Suppose that you find $\alpha$ dollars inside your envelope $A$.
%Now you are offered the options of keeping A or switching to B. What should you do? 
%\par \vskip0.3cm
%\par
%\noindent
%[(P1):Why is it paradoxical?].
%You reason that, with probability 1/2, the other envelope $B$ has either $\alpha/2$ or $2\alpha$ dollars. Thus the expected value (denoted $E_B(\alpha)$ at this moment) of the other envelope is $E_B(\alpha)=(1/2)(\alpha/2) + (1/2)(2\alpha)$ which is $1.25\alpha$. This is greater than the $\alpha$ in your current envelope $A$. Therefore,
%you should switch to B. But this seems clearly wrong, as your information about A and B is symmetrical. 
%This is the famous two-envelope paradox
%(i.e.,
%"The Other Person's Envelope is Always Greener"
%).
%%XXXXXXXXXXXXXXXXxxxxxxxxx
\par
\vskip0.5cm
\noindent
\par
Further consider the following problem, which is quite easy.
\par
\noindent
{\bf Problem 1$'$}
\rm
[The trivial two envelope problem].
The host presents you with a choice between two envelopes
(i.e.,
Envelope A
and
Envelope B).  
You know Envelope A [resp.
Envelope B]
includes 10 dollars
[resp.
20 dollars].
Define
$$ \overline{x} = 
\begin{cases}
20,\;\; (\mbox{ if } x=10),
\\
10,\;\; (\mbox{ if } x=20)
\end{cases}
$$
You choose randomly (by a fair coin toss) one envelope,
And you get
$x_1$ dollars
(i.e.,
if the envelope is A [resp B],
$x_1$ is equall to
10 dollars [resp. 20 dollars]).
And the host gets $\overline{x}_1$ dollars.
Next, by a similar way, you choose randomly (by a fair coin toss) one envelope,
And you get
$x_2$ dollars,
Repeating the trial, you get
$$
x_1, x_2, x_3, \cdots
$$
Then, you can assure, by the law of large numbers ({\it cf.} \cite{Ishi5, Ishi10, Kolm}), that
$$
\lim_{N \to \infty } \frac{x_1+ x_2 + \cdots + x_N}{N}
=
\lim_{N \to \infty } \frac{{\overline{x}}_1+ {\overline{x}}_2 + \cdots + {\overline{x}}_N}{N}=\frac{10+20}{2}
$$
Is it true?
Of course, it is true and not paradoxical.

\vskip0.2cm

\par
We consider that 
it is well known that
the above two problems
are essentially the same,
in spite that
the two are superficially different.
Thus, the purpose of this paper
is not to show the equality of the two problems,
but
to show that
\begin{itemize}
\item[$\bullet$]
the 
equivalence of two problems
(i.e.,
Problem 1 (the two-envelope paradox)
and
Problem 1$'$ (the trivial two-envelope paradox)
)
%(the St. Petersburg two-envelope paradox))
is automatically clarified,
if
Problems 1 is described in terms of quantum language.
\end{itemize}
%This will be done in Section 2. Thus, in the following section 1.2, according 
%to refs. {\cite{Ishi1}}-{\cite{Ishi100}}.
%we review quantum language.
%%if
%Problems 1 is described in terms of quantum language.
%\end{itemize}
This will be done in Section 3 (non-Bayesian two envelope paradox).
Also, we add
Section 4 (Bayesian two envelope paradox)
and
Section 5 (non-Bayesian St. Petersburg two envelope paradox).
In the following section 2, according 
to refs. {\cite{Ishi3}}-{\cite{Ishi100}}.
we review quantum language.

\vskip2.0cm

\section{Measurement theory
(= Quantum language)}

%\subsection{Review: classical quantum language}
\subsection{The motivation of quantum language}
In \cite{Merm}, N.D. Mermin introduced Feynman's two words
about quantum mechanics
as follows
({\it cf}. The Character of Physical Law (Cambridge: MIT Press, 1965)).
\begin{itemize}
\item[(A1)]
There was a time when the newspapers said that only twelve men understood the theory of relativity. I do not believe there ever was such a time. There might have been a time when only one man did, because he was the only guy who caught on, before he wrote his paper. But after people read the paper a lot of people understood the theory of relativity in some way or other, certainly more than twelve. On the other hand, I think I can safely say that {\it nobody understands quantum mechanics}.
\item[(A2)]
We have always had a great deal of difficulty understanding the world view that quantum mechanics represents.
$\cdots \cdots$
I cannot define the real problem, therefore I suspect there's no real problem, but I'm not sure {\it there's no real problem}.
\end{itemize}
For this significant Feynman's words, we assert that
\begin{itemize}
\item[(A3)]
If we start from the declaration
{\it "there's no real problem"},
it is a matter of course that {\it nobody understands quantum mechanics},
but
we can necessarily discover "quantum language",
which is located such as in Figure 1 (world views) below.
\end{itemize}
This is all my opinion concerning quantum language.
%\subsubsection{The classification and the location}
\par
\noindent
\par
%We think that the above problems have charm that beginners can challenge. 
%Thus, in consideration of undergraduate students,
%our concern is concentrated to
%"classical MT@(="CMT")
%in the following classification of measurement theory (="MT"="QL"="quantum language"):
%\begin{itemize}
%\item[(A4)]
%$
%%\quad
%\underset{\text{\scriptsize (=QL)}}{\text{MT}}
%$
%%$
%%\;
%%=
%$\left\{\begin{array}{ll}
%%\underset{\text{\scriptsize
%\underset{(\approx \text{"quantum mechanics"})}{\text{quantum MT }}
%&
%\begin{cases}
%\text{quantum PMT (pure measurement theory)}
%\\
%\text{quantum SMT (statistical measurement theory)}
%\end{cases}
%\\
%\\
%%\text{classical measurement theory,}
%%\text{classical MT}
%\underset{(\approx \text{"statistics"})}{\text{classical MT }}
%&
%\begin{cases}
%\text{classical PMT ($\approx$ non-Bayesian statistics)}
%\\
%\text{classical SMT($\approx$ Bayesian statistics)}
%\end{cases}
%\end{array}\right.
%$
%\end{itemize}
%That is, we are not concerned with quantum mechanics in this paper.
%Also,
%for the position of
%MT
%in science,
%see {\bf Figure 1}
%({\it cf.} {\cite{Ishi1}}-{\cite{Ishi100}}).
%%which was precisely explained in {}{\cite{Ishi5, Ishi8}}.

%Also, we assert that quantum language is located as follows:

\begin{center}
%\unitlength0.41mm
\begin{picture}(410,170)
%\thickness
%{
%\put(0,0){
%\thicklines
%\color{blue}
%\dashline[50]{4}(190,155)(110,155)(110,72)(400,72)(400,155)(290,155)}
%\color{black}
%}
%}
{
\put(10,70){
{
%\overset{\text{\scriptsize }}{
%\underset{\text{\scriptsize }}
\put(0,-3){
$\!\!
\underset{{\text{\footnotesize Alistotle}}}{\underset{{\text{\footnotesize Plato}}}{\overset{\text{\footnotesize Parmenides}}{{\overset{\text{\footnotesize Socrates}}{
{\fbox{\shortstack[l]{Greek\\ 
{\footnotesize philosophy}}}}
}
}
}
}
}
$
}
}
\put(51,-3){
\rm
$\xrightarrow[\text{\footnotesize sticism}]{\text{\footnotesize Schola-} }$
$\!\! \textcircled{\scriptsize 1}$
}
\put(93,7){
{\line(0,1){34}}
}
\put(93,-7){
{\line(0,-1){47}}
}
}
%%%%%%%%%%%%%%%%%%
%\thicklines
\put(100,70){
%$SSSSSSSSSSSSSSSSSSSSSSSSSSSSSSS
$
%\left\{
\begin{array}{l}
\!\!\!
{\; \xrightarrow[]{ \; 
\quad
%\tex
}}
\overset{\text{\scriptsize (monism)}}{\underset{\text{\scriptsize (realism)}}
{\fbox{\text{Newton}}}}
{
\overset{\textcircled{\scriptsize 2}}{{\line(1,0){17}}}
}
%\quad
%\xrightarrow[\qquad]{}
%%%%%%%%%%%%%%%%%%%
%%%%%%%%%%%%%%%%%%
%\left\{
\begin{array}{llll}
\!\!
\rightarrow
{\fbox{\shortstack[l]{relativity \\ theory}}}
{\xrightarrow[]{\qquad \quad \;\;\;}
}{\textcircled{\scriptsize 3}}
\\
\\
\!\!
\rightarrow
{\fbox{\shortstack[l]{quantum \\mechanics}}}
%%%%%%%%%%%%%
%%%%%%%%%%%%%%%%%
{
\xrightarrow[\qquad \quad \;\; ]{}
}\textcircled{\scriptsize 4}
%%%%%%%%%%%%%%%%%%%
\end{array}
%\right.
%%%%%%%%%%%%%%%%%
%%%%%%%%%%%%%%%%
\\
\\
%%%%%%%%%%%%%%%%%%%
%%%%%%%%%%%%%%%%%%%%
\!\! \xrightarrow[]{{
\quad}}
\overset{\text{\scriptsize (dualism)}}{
\underset{\text{\scriptsize (idealism)}}{\fbox{
{\shortstack[l]{Descartes\\ Rock,... \\Kant}}%%%
}}
}
{\xrightarrow[]{\textcircled{\scriptsize 6}
%\textcirc
}}%%
\!%\!\!
\overset{\text{\scriptsize (linguistic view)}}{\fbox{
\shortstack[l]{language \\ philosophy}%%%
}}
%%%%
\!\! \xrightarrow[{
}
{
\text{\footnotesize
%(axiomatization)
zation
}}
]{{
{
%\textcir
}%%%
{\text{\footnotesize
quanti-}
}
}}\textcircled{\scriptsize 8}
%%%%
%%%%%%%%%%%%%%%%%%%%%%
\end{array}
%\right.
$
}
%E
%%%%%%%%%%%%%%%%%%%%%%%%%%PPPPPPPPPPPPP
\put(300,86){
{\put(-40,0){\drawline(0,2)(0,-30)}}
{\put(-40,-32){\text{$\xrightarrow[]{\; \text{\footnotesize language}}$}}}
{\put(6,-32){\textcircled{\scriptsize 7}}}
}
\put(190,80){\line(0,1){46}}
%%%%%%%%%%%%%%%%%%%%%%%%%%%
\put(302,110){
$
\left.\begin{array}{llll}
\; 
\\
\; 
\\
\; 
\\
\;
\end{array}\right\}
\xrightarrow[]{\textcircled{\scriptsize 5}}
{\!\!\!
\overset{\text{\scriptsize (unsolved)}}{
\underset{\text{\scriptsize (quantum phys.)}}{
%\dashbox{2}(60,30)[c]
\fbox{\shortstack[l]{theory of \\ everything}}
}
}
}
$
}
%%%%%%%%%%%%%%%%%%%
\put(302,20){
$
\left.\begin{array}{lllll}
\; 
\\
\; 
\\
\; 
\\
\;
\\
\;
\end{array}\right\}
{\xrightarrow[]{\textcircled{\scriptsize 10}}}
%\dashbox{2}(80,30)[c]
\overset{\text{\scriptsize (=MT)}}{
\underset{\text{\scriptsize (language)}}{
\fbox{\shortstack[l]{\color{red}quantum\\ \color{red}language}}
}
}
$
}
%%%%%%%%%%%%%%%%%%%%%
%%%%%%%%%%%%%%%%%%%%%%%%
\put(100,-70){
{\bf 
%\color{magenta}
%\hypertarget{Pic1.1}{{}1.1}: 
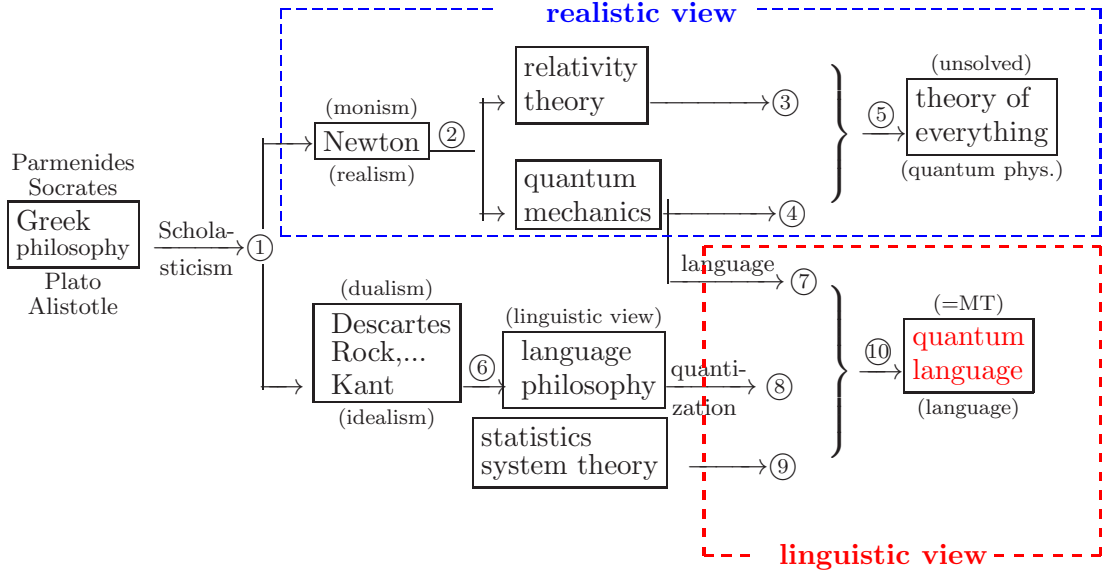
\hypertarget{Figure 1}{Figure 1}: The history of the world-view
}
}
\put(65,-32){
%For the explanations of
%\textcircled{\scriptsize 1}-\textcircled{\scriptsize 8},
%see [5,6].
%%{\bf
%%throughout this paper.}
}
{
\thicklines
\color{red}
%\allinethickness{0.9mm}
\dashline[50]{4}(287,-47)(270,-47)(270,70)(420,70)(420,-47)(380,-47)}
\thicklines
%{\put(208,-15)
{\put(175,-16)
%{\underset{\text{\scriptsize {}{}{}{}{}{}}}
{{
%\underset{\text{\scriptsize {}{}{}{}{}{}}}
{
%%\fbox{\shortstack[l]{{}{}{}}}}
\fbox{\shortstack[l]{ statistics \\ system theory}}
}
}$\xrightarrow[]{\qquad \;}$\textcircled{\scriptsize 9}
}
}
{
%\put(270,0){\dashbox{3}(190,70)[c]{}}
{
\put(288,-50){\color{red}
\bf
$\;\;$
linguistic view
%(A$_2$)
}
\color{black}
}
{
\put(200,155){\color{blue}
\bf
$\;\;$
realistic view
%(A$_2$)
}
\color{black}
}
}
}
%%\thickness
%%\put(270,0){\dashbox{3}(190,70)[c]{}}
%{
%\put(110.155){\color{blue}
%\bf
%$\;\;$
%}
%}
%{
%\put(0,0){
%\thicklines
{
\color{blue}
$\!\!\!\!\!\!\!\!${\dashline[50]{4}(190,160)(110,160)(110,74)(420,74)(420,160)(290,160)}
%}
\color{black}
}
\end{picture}
\vskip1.8cm
\end{center}

\vskip1.5cm

\subsection{The classifications of measurement theory}

\rm
\par
\par
\noindent
\par
In this section,
we introduce measurement theory (or in short, MT). This theory is a kind of language,
and thus, it is also called quantum language
(or in short, QL).

%\newpage
\par
\par
\rm
Measurement theory
({\it cf.} refs.\textcolor{black}{\cite{Ishi3}-\cite{Ishi100}})
is,
by an analogy of
quantum mechanics
(or,
as a linguistic turn of quantum
mechanics
), constructed
as the scientific
theory
%physical language})
%%[${}{}1\text{--}7}${
formulated
in a certain 
%\linebreak
{}{$C^*$}-algebra ${\cal A}$
(i.e.,
a norm closed subalgebra
in the operator algebra $B(H)$
composed of all bounded linear operators on a Hilbert space $H$,
{\it cf.$\;$}\textcolor{black}{\cite{Neum, Saka}}
). 

When ${\cal A}=B_c(H)$,
the ${C^*}$-algebra composed
of all compact operators on a Hilbert space $H$,
the MT is called {quantum measurement theory}
(or,
quantum system theory),
which can be regarded as
the linguistic aspect of quantum mechanics.
Also, when ${\cal A}$ is commutative
$\big($
that is, 
when ${\cal A}$ is characterized by $C_0(\Omega)$,
the $C^*$-algebra composed of all continuous 
complex-valued functions vanishing at infinity
on a locally compact Hausdorff space $\Omega$
({\it cf.$\;$}\textcolor{black}{\cite{Saka}})$\big)$,
the MT is called {classical measurement theory}.
Thus, we have the following classification:
\begin{itemize}
\item[(B1)]
$
\underset{\text{\scriptsize }}{\text{MT}}
$
%$
%\;
%=
$\left\{\begin{array}{ll}
%\underset{\text{\scriptsize
\text{quantum MT$\quad$(when non-commutative ${\cal A}=B_c (H)$
)
%,
%$L^\infty (\Omega )=B(H)$)
}
%}
%
\\
\\
%\\
%\text{classical measurement theory,}
\text{classical MT
%\underset{\text{\scriptsize 
$\quad$
(when commutative ${\cal A}=C_0(\Omega)$
%,
%$L^\infty (\Omega )=L^\infty (\Omega, \nu)(\subseteq B(L^2(\Omega, \nu) )$
)}
\end{array}\right.
$
\end{itemize}
In this paper, we devote ourselves to
CMT(= classical MT), which is classified as follows.
\begin{itemize}
\item[(B2)]
$
\underset{\text{\scriptsize }}{\text{CMT}}
$
%$
%\;
%=
$\left\{\begin{array}{ll}
%\underset{\text{\scriptsize
\text{(B2$_1$): PCMT
(= pure classical measurement theory)
in $\S$2.3}
%}
%
\\
\\
%\\
%\text{classical measurement theory,}
\text{(B2$_2$): SCMT
(= statistical classical measurement theory)
in $\S$2.4}
%
%
%\text{classical MT
%%\underset{\text{\scriptsize 
%$\quad$
%(when commutative ${\cal A}=C_0(\Omega)$),
%$L^\infty (\Omega )=L^\infty (\Omega, \nu)(\subseteq B(L^2(\Omega, \nu) )$)}
%%
%%
\end{array}\right.
$
\end{itemize}

\vskip2.5cm

\subsection{The preparation of CMT}
\par
\noindent
\par
For the general theory of measurement theory, see refs. \cite{Ishi3}-\cite{Ishi100}.
In order to read this paper,
it suffices to know the following.

\par
\noindent
\par
Let $\Omega$ be a locally compact space.
%, which is called a {\it state space}.
Define the continuous functions space $C_0(\Omega)$ such that
$$
C_0(\Omega)
=
\{
f
\;|\;
\mbox{
$f$ is a complex valued continuous function on $\Omega$
such that
$\lim_{\omega \to \infty} f(\omega)=0$
}
\}
$$
which is a Banach space (or precisely,
commutative $C^*$-algebra
)
with the norm
$\| f \| = \max_{\omega \in \Omega} |f(\omega )|$.
\par
Let
${\mathcal M}(\Omega)$ be the dual Banach space of
$C_0(\Omega)$,
i.e., ${\mathcal M}(\Omega)=C_0(\Omega)^*$.
Riese theorem ({\it cf.} ref. \cite{Yosi}) says that
${\mathcal M}(\Omega)$ is the space of all finite complex-valued measures
on $\Omega$.
Thus, we denote that
$$
{}_{C_0(\Omega)^*} \big\langle
\rho,
f \big\rangle{}_{C_0(\Omega )}=
\int_\Omega f(\omega ) \rho (d \omega )
%f(\omega)
\qquad
(\forall f \in C_0(\Omega),
\quad
\forall \rho \in {\mathcal M}(\Omega)=C_0(\Omega)^*
)
$$
The
${\mathcal M}^m(\Omega)$
(i.e.,
the space of all probability measures on
$\Omega$
)
is called a
{\it
mixed state class}.
An element $\rho (\in {\mathcal M}^m(\Omega))$
is called a {\it mixed state}.
%(or in short, {\it state}).
%The element $\omega ( \in \Omega )$
%is called a {\it pure state} (or in short, {\it state}).

For each $\omega \in \Omega$,
define the point measure $\delta_\omega (\in {\mathcal M}^m(\Omega))$ such that
$$
{}_{C_0(\Omega)^*}
\big\langle
\delta_\omega,
f \big\rangle{}_{C_0(\Omega )}=f(\omega)
\qquad
(\forall f \in C_0(\Omega))
$$
The
${\mathcal M}^p(\Omega)$
(i.e.,
the space of all point measures on
$\Omega$
)
is called a 
{\it pure state class}.
An element $\rho (\in {\mathcal M}^p(\Omega))$
is called a {\it pure state}
(or in short, {\it state}).
Under the identification:
${\mathcal M}^p(\Omega) \ni \delta_\omega \longleftrightarrow \omega \in \Omega$,
the $\omega (\in \Omega)$ is also called a {\it state}
(or precisely,
{\it pure state}). 

Let $\nu$ be a fixed ($\sigma$-finite) measure on $\Omega$
such that
$$
\nu ( K ) < \infty, 0 < \nu(D)
\qquad
(\forall \mbox{ compact set $K$},
\forall \mbox{ open set $D$}
( \in {\mathcal B}_{ \Omega}:
\mbox{the Borel field in $\Omega$} )
%}
)
$$
Define the Banach space $L^p(\Omega, \nu )$ $(1 \le p \le \infty)$ such that
$$
f \in
L^p(\Omega )
\Leftrightarrow
\mbox{
$f$ is a complex-valued measurable function on $\Omega$
such that
$\|f \|_{L^p(\Omega)}< \infty $
}
$$
where
$\|f \|_{L^p(\Omega)}=\Big[ \int_\Omega |f(\omega |^p \nu (d \omega )\Big]^{1/p}$
$(1 \le p <\infty)$,
$= \inf \{ a \ge 0 : \nu(\{ \omega\;:\; |f(\omega )| > a\})=0 \}$
$(p= \infty )$.
%
%Now we shall explain the measurement theory
%(i.e.,
%classical SMT${}^{W^*}$
%).
%\par
%\noindent
%\par
%Let
%$[{C_0(\Omega )} \subseteq L^\infty (\Omega ) {}]$
%be the fundamental structure
%of measurement theory.
%Let
%${L^\infty (\Omega )}_*$ be the
%pre-dual Banach space of
%${L^\infty (\Omega )}$.
%That is,
%$ {L^\infty (\Omega )}_* $
%$ {=}  $
%$ \{ \rho \; | \; \rho$
%%: L^\infty (\Omega ) \to {\mathbb C} $
%is a weak$^*$ continuous linear functional on ${L^\infty (\Omega )}$
%$\}$,
%and
%the norm $\| \rho \|_{ {L^\infty (\Omega )}_* } $
%is defined by
%$ \sup \{ | \rho ({}F{}) |  \:{}: \; F \in {L^\infty (\Omega )}
%\text{ such that }\| F \|_{L^\infty (\Omega )} 
%(=\| F \|_{B(H)} )\le 1 \}$.
%The bi-linear functional
%$\rho(F)$
%is
%also denoted by
%${}_{{L^\infty (\Omega )}_*}
%\langle \rho, F \rangle_{L^\infty (\Omega )}$,
%or in short
%$
%\langle \rho, F \rangle$.
%Define the
%\it
%mixed state
%$\rho \;(\in{L^\infty (\Omega )}_*)$
%\rm
%such that
%$\| \rho \|_{L^\infty (\Omega )_* } =1$
%and
%$
%\rho ({}F) \ge 0
%\text{ 
%for all }F\in {L^\infty (\Omega )}
%\text{ satisfying }
%F \ge 0$.
%And put
%\begin{align*} {\frak S}^m  ({}{L^\infty (\Omega )}_*{})
%{=}
%\{ \rho \in {L^\infty (\Omega )}_*  \; | \;
%\rho
%\text{ is a mixed state}
%\}.
%\end{align*}
%%
%%\rm
%\rm
%
%
%According to the noted idea ({\it cf.} ref. \textcolor{black}{\cite{ Davi}})
%in quantum mechanics,

Motivated by a nice idea in
ref.\cite{Davi},
an {\it observable}
${\mathsf O}{\; \equiv}(X, {\cal F},$
$F)$ in the 
$L^\infty (\Omega, \nu  )$
is defined as follows:
%, if it satisfies:
\par
\par
\begin{itemize}
\item[(C1)]
[$\sigma$-field]
$X$ is a set,
${\cal F}
(\subseteq 2^X $,
the power set of $X$)
is a $\sigma$-field of $X$,
that is,
{\lq\lq}$\Xi_1, \Xi_2, \Xi_3, \cdots \in {\cal F}\Rightarrow \cup_{k=1}^\infty \Xi_k \in {\cal F}${\rq\rq},
{\lq\lq}$X \in {\mathcal F}${\rq\rq}
and
{\lq\lq}$\Xi  \in {\cal F}\Rightarrow X \setminus \Xi \in {\cal F}${\rq\rq}.
\item[(C2)]
[Countably additivity]
$F$ is a mapping from ${\cal F}$ to ${{L^\infty (\Omega, \nu  )}}$ 
satisfying:
%\begin{enumerate}
%\item
(a):
for every $\Xi \in {\cal F}$, $F(\Xi)$ is a non-negative element in 
$L^\infty (\Omega, \nu  )$
such that $0 \le F(\Xi) $
$\le I$, 
(b):
$F(\emptyset) = 0$ and 
$F(X) = I$,
%($\forall \rho \in {\frak S}^p  ({}L^1(\Omega ){})$),
where
$0$ and $I$ is the $0$-element and the identity
in $L^\infty (\Omega, \nu  )$
respectively.
(c):
for any countable decomposition $\{ \Xi_1,\Xi_2, \ldots \}$
of $\Xi$
$\in {\cal F}$
(i.e., $\Xi_k , \Xi \in {\cal F}$
such that
$\bigcup_{k=1}^\infty \Xi_k = \Xi$,
$\Xi_i \cap \Xi_j= \emptyset
(i \not= j)$),
it holds that
\begin{align}
\quad
\lim_{K \to \infty } 
{}_{{}_{L^1(\Omega )}}
\langle \rho, 
F( \bigcup_{k=1}^K \Xi_k )
 \rangle_{{}_{L^\infty (\Omega, \nu  )}}
=
&
{}_{{}_{L^1(\Omega )}}
\langle \rho, 
 F( \Xi ) 
 \rangle_{{}_{L^\infty (\Omega, \nu  )}}
\Big(
\equiv
\int_\Omega
\rho(\omega) \cdot [F(\Xi)](\omega)
\nu(d \omega )
\Big)
\label{eq2}
\\
&
%\rangle_{L^\infty (\Omega )}
\quad
\big(
\forall \rho \in L^1(\Omega, \nu )
\big)
\nonumber
\end{align}
i.e.,
$
\lim_{K \to \infty }  
F( \bigcup_{k=1}^K \Xi_k )
= F( \Xi ) 
$
in the sense of weak${}^*$ convergence in ${L^\infty (\Omega, \nu  )}$.
\end{itemize}

\par
\noindent
%As the simplest observable, the {\it existence observable}
%${\mathsf O}^{\rm (e)}{=}(X, \{\emptyset, X \},$
%$F^{\rm (e)})${L^\infty (\Omega )}_*
%is defined such that
%$F^{\rm (e)}(\emptyset )=0$
%and
%$F^{\rm (e)}(X )=I$.
%
%
\par
\vskip0.3cm
\par
\par
\noindent
%{\it Re
%By the Hopf extension theorem
%({\it cf.$\;$}\textcolor{black}{\cite{Yosi}}),
%we have the mathematical probability space
%$(X,$
%$ {\overline{\cal F}},$
%$ \rho^m (F(\cdot )) \;)$ 
%where
%${\overline{\cal F}}$
%is the smallest $\sigma$-field such that
%${{\cal F}} \subseteq {\overline{\cal F}}$.
%For the other formulation
%(i.e.,
%$W^*$-algebraic formulation
%),
%see
%the appendix in \textcolor{black}{\cite{Ishi6}}.
%

\par
\noindent
Let $\{f_n\}_{n=1}^\infty$ be a sequence in $L^1(\Omega, \nu )$.
And let $\rho_0 \in {\mathcal M}^m (\Omega )$.
Here,
"$w^*-\lim_{n \to \infty } f_n = \rho_0$" means that
$$
\lim_{n \to \infty }
\int_\Omega f_n(\omega ) \cdot \phi (\omega ) \nu(d \omega )
=
\int \phi(\omega ) \rho_0(d \omega )
\quad
(\forall \phi \in C_0(\Omega )  )
$$
And,
we say that
{\it
"$F (\in L^\infty ( \Omega , \nu ) )$
is essentially continuous at $\rho_0 (\in {\mathcal M}^m (\Omega ))$",}
if there uniquely exists a complex number $\gamma$ such that
$$
"w^*-\lim_{n \to \infty } f_n = \rho_0"
\Longrightarrow
"\lim_{n \to \infty }
\int_\Omega F(\omega ) \cdot f_n(\omega ) \nu ( d \omega ) =
\gamma"
$$
And we denote that
$
%\sset{{\
\rho_0 ( F )
%= \int_\Xi F(dx)(\rho) \big).
$
$($
$=
{}_{{C_0(\Omega)^*}}\langle
\rho_0,
F
\rangle_{{L^\infty (\Omega, \nu )}}
)
=
\gamma$.
\vskip1.0cm
\par
\noindent
\bf
Remark 1.
\rm
Without loss of generality,
we can assume that
$\Omega$ is compact,
and $\nu(\Omega)=1$.

\vskip1.0cm

\par
\subsection{Pure Classical Measurement Theory}
\par
\noindent

\par
%Now
%we shall explain PMT
%in
%(A$_1$),
%
%Our concern in this paper is SMT in (A$_1$),
%which is constructed on the base of PMT.
%Thus, we begin with PMT.
%%22 22 22 2.2 2.2 2.2
\par
\rm

With any {\it classical system} $S$, a fundamental structure
$[{C_0(\Omega )} \subseteq L^\infty (\Omega, \nu  ) {}]$
can be associated in which the 
pure
measurement theory (B2$_1$) of that system can be formulated.
A {\it pure state} of the system $S$ is represented by an element
$\delta_\omega (\in {\mathcal M}^p  (\Omega)$="pure state class"({\it cf.} ref.\cite{Ishi6}))
and an {\it observable} is represented by an observable 
${\mathsf{O}}{\; =} (X, {\cal F}, F)$ in ${{L^\infty (\Omega, \nu  )}}$.
Also, the {\it measurement of the observable ${\mathsf{O}}$ for the system 
$S$ with the pure state $\delta_\omega$}
is denoted by 
${\mathsf{M}}_{{{L^\infty (\Omega, \nu  )}}} ({\mathsf{O}}, S_{[\delta_\omega]})$
$\big($
or more precisely,
${\mathsf{M}}_{L^\infty (\Omega, \nu  )} ({\mathsf{O}}{\; =} (X, {\cal F}, F), S_{[\delta_\omega]})$
$\big)$.
An observer can obtain a measured value $x $
($\in X$) by the measurement 
${\mathsf{M}}_{L^\infty (\Omega, \nu )} ({\mathsf{O}}, S_{[\delta_\omega]})$.
\par
\noindent
%%\vskip0.1cm
\par
The Axiom$^{\rm \small PCMT}$\;1 presented below is 
a kind of mathematical generalization of Born's probabilistic interpretation of quantum mechanics.
%And thus, it is a statement without reality.
\par
\noindent
%$\bf{Axiom \;1}$\rho^p
{\bf{Axiom$^{\rm \small PCMT}$\;1
\rm
$[$Pure Measurement$]$}}.
\it
The probability that a measured value $x$
$( \in X)$ obtained by the measurement 
${\mathsf{M}}_{{{L^\infty (\Omega, \nu  )}}} ({\mathsf{O}}$
${ \equiv} (X, {\cal F}, F),$
{}{$ S_{[\delta_{\omega_0}]})$}
%$\big)$.
%
%${\mathsf{M}}_{{{L^\infty (\Omega )}}} ({\mathsf{O}}, S_{[\rho^p_0]})$ 
belongs to a set 
$\Xi (\in {\cal F})$ is given by
$
%\sset{{\
\delta_{\omega_0}( F(\Xi))
%= \int_\Xi F(dx)(\rho^p) \big).
$
$\Big(=
{}_{{C_0(\Omega)^*}}\langle
\delta_{\omega_0},
F(\Xi)
\rangle_{{L^\infty (\Omega, \nu )}}
\Big)$,
if $F(\Xi)$ is essentially continuous at $\delta_{\omega_0}$.
%({\it cf.}{\rm ref.\cite{Ishi6}}).
\rm

\par
\vskip0.5cm
\par
\noindent
\par
Let $[{C_0(\Omega, \nu )} \subseteq L^\infty (\Omega, \nu ) {}]$
be a fundamental structure.
We shall introduce the following notation:
\rm
It is usual to consider that
we do not know the pure state
$\delta_\omega$
$(
\in
{\mathcal M}^p  (\Omega)
)$
when
we take a measurement
${\mathsf{M}}_{{{L^\infty (\Omega, \nu )}}} ({\mathsf{O}}, S_{[\delta_\omega]})$.
That is because
we usually take a measurement ${\mathsf{M}}_{{{L^\infty (\Omega, \nu )}}} ({\mathsf{O}},
S_{[\delta_\omega]})$
in order to know the state $\delta_\omega$.
Thus,
\begin{itemize}
\item[(D1)]
when we want to emphasize that
we do not know the state $\delta_\omega$,
${\mathsf{M}}_{{{L^\infty (\Omega, \nu )}}} ({\mathsf{O}}, S_{[\delta_\omega]})$
is denoted by
${\mathsf{M}}_{{{L^\infty (\Omega, \nu )}}} ({\mathsf{O}}, S_{[\ast]})$
\item[(D2)]
also,
when we know the distribution $\rho_0$
$( \in {\mathcal M}^m(\Omega) )$
of the unknown state
$\delta_\omega$,
the
${\mathsf{M}}_{{{L^\infty (\Omega, \nu )}}} ({\mathsf{O}},$
$ S_{[\delta_\omega]})$
is denoted by
${\mathsf{M}}_{{{L^\infty (\Omega, \nu )}}} ({\mathsf{O}}, S_{[\ast]}
({ \rho_0}) )$.
The $\rho_0$
is called a mixed state.
\end{itemize}
%({\it cf}. \cite).

We have the following fundamental theorem
in measurement theory:

\par
\noindent
{\bf Theorem 1}
[Fisher's maximum likelihood method
({\rm cf}. {}{\cite{Ishi4}-\cite{Ishi11}})].
Assume that
a measured value $x (\in X)$ is obtained
by a measurement
${\mathsf M}_{L^\infty(\Omega,\nu)}({\mathsf O}:=(X,{\cal F}, F) , S_{[*]})$.
Put
\begin{align}
%%%\label{3fmlm_cor}
f(x, \omega)=
\inf_{\omega_1 \in \Omega }
\Big[\lim_{\Xi \to \{ x \}, \Xi \ni x, [{F}(\Xi )](\omega_1) \not= 0} \frac{[{F}(\Xi )](\omega)}{
[{F}(\Xi )](\omega_1)
}
\Big]
%
%\inf_{\omega_1 \in \Omega }
%\Big[\lim_{{\mathcal F} \ni \Xi \to \{ x \}, [{F}(\Xi )](\omega_1) \not= 0} \frac{[{F}(\Xi )](\omega)}{
%[{F}(\Xi )](\omega_1)
%}
%\Big]
\qquad
(\forall \omega \in \Omega )
\label{eq3}
\end{align}
Then,
there is a reason to infer that
the unknown state
$[\ast]$
is equal to 
$\delta_{\omega_0} \;(\in \Omega )$
such that
%which
%K \subseteq {\frak S}^m ({C_0(\Omega)}^*))$
%%satisfies
\begin{align*}
%%%\label{3fmlm_cor}
f(x, \omega_0)=1
\end{align*}
Also,
if $f(x, \omega_1) =0$,
then
there is no possibility that
$[\ast]$
$=$
$\delta_{\omega_1}$.

%\rm
%\par
%%For the further argument
%%(i.e.,
%%the $W^*$-algebraic formulation) of measurement theory,
%%see
%Appendix in \textcolor{black}{\cite{Ishi7}}.
%\noindent
%$\b
%Note that
\par
\vskip0.2cm
\par
\noindent
\par

\par
\noindent
\bf
Definition 1
\rm
[Parallel measurement].
Consider two measurements:
${\mathsf M}_{L^\infty(\Omega_1, \nu_1)}({\mathsf O}_1:=(X_1,{\cal F}_1, F_1) ,$
$ S_{[\delta_{\omega_1}]})$
and
${\mathsf M}_{L^\infty(\Omega_2, \nu_2)}({\mathsf O}_2:=(X_2,{\cal F}_2, F_2) , S_{[\delta_{\omega_2}]})$.
Let
$(\Omega_1 \times \Omega_2, \nu_1\otimes \nu_2)$
be the product measure space
of
$(\Omega_1, \nu_1)$
and
$(\Omega_2, \nu_2)$.
And consider the parallel measurement
${\mathsf M}_{L^\infty(\Omega_1 \times \Omega_2, \nu_1\otimes \nu_2)}$
$({\mathsf O}_1 \otimes {\mathsf O}_2:=(X_1 \times X_2,
{\cal F}_1 \boxtimes {\cal F}_2
, F_1 \otimes F_2) , S_{[\delta_{(\omega_1, \omega_2)}]})$,
which is denoted by
$\bigotimes_{n=1}^2
{\mathsf M}_{L^\infty(\Omega_n, \nu_n)}({\mathsf O}_n:=(X_1,{\cal F}_n, F_n) ,$
$ S_{[\delta_{\omega_n}]})$.
Here, ${\cal F}_1 \boxtimes {\cal F}_2$ is the product field of
$
{\cal F}_1$
and
${\cal F}_2
$. And, $
F_1 \otimes F_2
$
is defined by
$$
[(F_1 \otimes F_2)(\Xi_1 \times \Xi_2 )](\omega_1, \omega_2 )
=
[F_1(\Xi_1)](\omega_1)
\cdot
[F_2(\Xi_2)](\omega_2)
$$
($\forall (\omega_1, \omega_2 ) \in \Omega_1 \times \Omega_2$,
$\forall \Xi_1 \in {\mathcal F}_1$,
$\forall \Xi_2 \in {\mathcal F}_2$
).
Here,
the linguistic interpretation of quantum mechanics  ({\it cf.} \cite{Ishi6, Ishi7} ) asserts the identification:
"(D3)+(D4)"
$\Leftrightarrow$
"(D5)":
\begin{itemize}
\item[(D3)] a measured value $x_1 (\in X_1)$
is obtained by a measurement ${\mathsf M}_{L^\infty(\Omega_1, \nu_1)}({\mathsf O}_1:=(X_1,{\cal F}_1, F_1) , S_{[\delta_{\omega_1}]})$
\item[(D4)] a measured value $x_2 (\in X_2)$
is obtained by a measurement ${\mathsf M}_{L^\infty(\Omega_2, \nu_1)}({\mathsf O}_2:=(X_2,{\cal F}_2, F_2) , S_{[\delta_{\omega_2}]})$
\item[(D5)] a measured value $(x_1 ,x_2) (\in X_1 \times X_2)$
is obtained by a measurement ${\mathsf M}_{L^\infty(\Omega_1 \times \Omega_2\nu_1\otimes \nu_2)}({\mathsf O}_1 \otimes {\mathsf O}_2:=(X_1 \times X_2,
{\cal F}_1 \boxtimes {\cal F}_2
, F_1 \otimes F_2) , S_{[\delta_{(\omega_1, \omega_2)}]})$
\end{itemize}

%%_1C_0C_0
%\subsec

\par
\vskip0.5cm
\noindent
\par
This definition is generalized as follows.
For each
$n \in {\mathbb N}=\{1,2, \cdots \}$,
consider a measurement:
${\mathsf M}_{L^\infty(\Omega_n, \nu_n)}({\mathsf O}_n:=(X_n,{\cal F}_n, F_n) ,$
$ S_{[\delta_{\omega_n}]})$.
Let
$(\bigtimes_{n \in {\mathbb N}}
\Omega_n , \bigotimes_{n \in {\mathbb N}} \nu_n )$
be the infinite product probability measure space
({\it cf.} Remark 1 and \cite{Kolm}).
Let
$(\bigtimes_{n \in {\mathbb N}}
X_n , {\large \boxtimes}_{n \in {\mathbb N}}
{\mathcal F}_n )$
be the infinite product measurable space.
Then,
we have the parallel observable
$\bigotimes_{n \in {\mathbb N}}{\mathsf O}_n$
$=$
$(\bigtimes_{n \in {\mathbb N}}
X_n , {\large \boxtimes}_{n \in {\mathbb N}}{\mathcal F}_n,
\bigotimes_{n \in {\mathbb N}}F_n
)$
in
$L^\infty (
\bigtimes_{n \in {\mathbb N}}
\Omega_n , \bigotimes_{n \in {\mathbb N}} \nu_n )
)$
such that
$$
[(\bigotimes_{n \in {\mathbb N}}F_n)(\bigtimes_{n \in {\mathbb N}}\Xi_n
)](\omega_1, \omega_2, \cdots )
=
\bigtimes_{n \in {\mathbb N}}
[F_n(\Xi_n)](\omega_n)
\quad
(\forall \Xi_n \in {\mathcal F_n},
\forall (\omega_1, \omega_2, \cdots) \in \bigtimes_{n \in {\mathbb N}}
\Omega_n
)
$$
where a set $\{n \in {\mathbb N} \;|; \Xi_n \not= X_n \}$
is finite.
\par
\noindent
Thus, we have the infinite parallel measurement:
$\bigotimes_{n=1}^\infty
{\mathsf M}_{L^\infty(\Omega_n, \nu_n)}({\mathsf O}_n:=(X_1,{\cal F}_n, F_n) ,$
$ S_{[\delta_{\omega_n}]})$,
i.e.,
\begin{align}
{\mathsf M}_{L^\infty(
\bigtimes_{n \in {\mathbb N}}
\Omega_n , {\bigotimes}_{n \in {\mathbb N}} \nu_n
)}(
\bigotimes_{n \in {\mathbb N}}{\mathsf O}_n=
({\Large \bigtimes}_{n \in {\mathbb N}}
X_n , {\Large \boxtimes}_{n \in {\mathbb N}}{\mathcal F}_n,
\otimes_{n \in {\mathbb N}}F_n
),
S_{[\delta_{(\omega_1,\omega_2, \cdots )}]})
\label{eq4}
\end{align}

\noindent
%\vskip0.5cm
%\par
%\noindent
%{\bf
%2.2.
%Interpretation}\rho^p\rho
%\par
\par
\noindent
\vskip1.5cm
\par
%\noindent
%indent

%(i.e., Saussure's linguistic world view).
%
%
\noindent
\noindent
\subsection{Statistical Classical Measurement Theory
}

\par
%\vskip0.3cm

\par
\noindent
\par
The Axiom$^{\rm \small SCMT}$\;1 presented below 
is also
a kind of mathematical generalization of Born's probabilistic interpretation of quantum mechanics.

\par

\par
\noindent
%$\bf{Axiom \;1}$
{\bf{Axiom$^{\rm \small SCMT}$\;1\;
\rm
\;[Statistical measurement]}}.
\it
Recall the (D2).
The probability that a measured value $x$
$( \in X)$ obtained by the measurement 
${\mathsf{M}}_{{{L^\infty (\Omega, \nu )}}} ({\mathsf{O}}$
${ \equiv} (X, {\cal F}, F),$
{}{$ S_{[\ast]}({ \rho_0 }) )$}
%$\big)$.
%
%${\mathsf{M}}_{{{L^\infty (\Omega, \nu )}}} ({\mathsf{O}}, S_{[\rho_0]})$ 
belongs to a set 
$\Xi (\in {\cal F})$ is given by
$
%\sset{{\
\rho_0 ( F(\Xi) )
%= \int_\Xi F(dx)(\rho) \big).
$
$($
$=
{}_{{C_0(\Omega)^*}}\langle
\rho_0,
F(\Xi)
\rangle_{{L^\infty (\Omega, \nu )}}$
$)$
if $F(\Xi)$ is essentially continuous at $\rho_0$.
%({\it cf.} ref.\cite{Ishi7}).
\rm

\vskip0.4cm

\par
\noindent
{\bf Theorem 2}
[Bayes' method
({\rm cf}. {}{\cite{Ishi3}-\cite{Ishi11}})].
Assume that
a measured value $x (\in X)$ is obtained
by a measurement
${\mathsf M}_{L^\infty (\Omega, \nu)}({\mathsf O}:=(X,{\cal F}, F) , S_{[*]}(\rho_0))$.
Thus,
we can assert that:
\begin{itemize}
\item[(E)]
When
we
know 
a measured value
$x (\in X)$
obtained by
a
statistical
measurement
${\mathsf M}_{L^\infty(\Omega, \nu)}(
{\mathsf O} \equiv (X, {\cal F}, F)
,$
$ S_{[*]}
{(\rho_0 )}
)$,
there is a reason to
infer
that
the post-state
(i.e.,
the mixed state after the measurement
)
is
equal to
$\rho_{\mbox{\scriptsize post}}^x$
$( \in {\mathcal M}^m (\Omega, \nu ))$,
where
\begin{align}
\rho_{\mbox{\scriptsize post}}^x
= \lim_{\Xi \to \{x\} }\frac{ [F(\Xi)](\omega)
\;
\rho_0  }{
\int_{\Omega} [F(\Xi)](\omega)
\;
\rho_0 (d \omega ) 
\;}
%= [G( \Gamma)](\omega_0)
%\quad
%(\forall D \in {\mathcal B}_\Omega ).
\label{eq5}
\end{align}
\end{itemize}
\vskip1.0cm

\par
\noindent
\bf
Remark 2
\rm
[Bayesian statistics].
When
Bayes' theorem is used in SCMT,
SCMT
is called Bayesian statistics.
In Bayesian statistics, the mixed state $\rho_0$
may be called a "pretest state"
({\it cf.} refs.\textcolor{black}{\cite{Ishi4}--\cite{Ishi11}}
).
\par
\vskip0.5cm
\par
%\noindent
%\bf
%Remark 2.
%\rm
%[$C^*$-algebraic formulation and
%$W^*$-algebraic formulation
%].
%As mentioned in {\cite{Ishi2}-\cite{Ishi100}},
%quantum language has two formulation
%(i.e., $C^*$-algebraic formulation
%and
%$W^*$-algebraic formulation
%).
%Let ${\mathsf O} \equiv (X, {\cal F}, F)$ be an observable.
%In $C^*$-algebraic [resp.
%$W^*$-algebraic
%]
%formulation, it holds that
%$F(\Xi )$ 
%$\in C_0(\Omega )$
%[resp.
%$\in L^\infty (\Omega, \nu )$
%].
%In this paper,
%we devote ourselves to the
%$W^*$-algebraic formulation.
%For the
%$C^*$-algebraic approach to
%the two-envelope paradox,
%see
%\cite{Ishi13}.
%Now we consider that
%the
%$W^*$-algebraic approach
%is superior to the
%$C^*$-algebraic approach
%as far as the two-envelope problem is concerned.

\par
\vskip0.5cm
\par
\noindent
\bf
Remark 3
\rm
[Overview; quantum language
({\rm cf}. {}{\cite{Ishi3}-\cite{Ishi100}})
].
Although, in order to read this paper, it suffices to understand Axiom 1
(measurement:
Axiom$^{\rm \small PCMT}$\;1
and
Axiom$^{\rm \small SCMT}$\;1
),
we want to mention the overview of quantum language as follows.
Quantum language (=QL) is a kind of metaphysics
(i.e.,
language)
that has the following structure:
$$
\fbox{\mbox{QL}}
=
\overset{\mbox{(Axiom 1)}}{
\fbox{\mbox{measurement}}
}
+
\overset{\mbox{(Axiom 2)}}{
\fbox{\mbox{causality}}
}
+
\overset{\mbox{(the manual how to use Axioms 1 and 2)}}{
\fbox{\mbox{Linguistic interpretation}}
}
$$
And quantum language says that
\begin{itemize}
\item[(F1)]
Follow examples of the wordings in Axioms 1 and 2,
and
describe every phenomenon!
\end{itemize}
Applying a trial-and-error method
repeatedly, you may make progress
without the manual
(i.e.,
the linguistic interpretation).
In fact, the author has mastered the linguistic interpretation now at last by the trial and error for about twenty years. 
In this sense,
the manual
(i.e.,
the linguistic interpretation)
is
not absolutely indispensable for quantum language.
That is, we consider that
the term
"interpretation"
should not exist in physics but in metaphysics.
However,
it is earlier for progress to know the manual. 
For example, the following two:
\begin{itemize}
\item[(F2)]
consider the dualism (i.e.,
observer and object)
\item[(F3)]
only one measurement is permitted
\end{itemize}
are leading indicators.
Although we believe that
the linguistic interpretation should be determined uniquely
and naturally,
this is not guaranteed.
However,
if it is not determined uniquely,
it suffices to
add something to axioms.
This is the fate of metaphysics.
Also, note that
\begin{itemize}
\item[(F4)]
the competitor of quantum language (i.e.,
the linguistic interpretation of quantum mechanics) is statistics and is not physics
(i.e.,
the several interpretations of quantum mechanics).
\end{itemize}
Of course, we believe that
quantum language is forever.
%Thus, we are optimistic.
%
%
%\par

%3333333333333333333333
\vskip2.5cm
\section{Non-Bayesian approach to the two envelopes problem}
\par
\vskip1.0cm
%\noindent
%\subsection{(P1):The two-envelope paradox in Classical PMT ($\approx$ non-Bayesian statistics)}
%The quantum linguistic description in non-Bayesian statistics}
\subsection{The simple answer in which it is hard to notice the mistake (\ref{eq1})}
%Let $x$ be an unknown fixed positive number. 
\par
\noindent
\par
Consider the classical fundamental structure such that
$$
[C_0 (\Omega )
\subseteq
L^\infty ( \Omega, \nu )
%\subseteq
%B(L^2(\Omega, \nu ))
]
$$
Put $X=\overline{\mathbb R}_+ =\{ x \;|\;\mbox{$x$ is a non-negative real number}
\}$.
Let $V_1: \Omega \to \overline{\mathbb R}_+$ and
$V_2: \Omega \to \overline{\mathbb R}_+$
be continuous maps.
You may think that $V_2(\omega ) = 2 V_1 (\omega) \;\; (\forall \omega \in \Omega )$.

For each $k=1,2$,
define the observable 
${\mathsf O}_k=(X(=\overline{\mathbb R}_+),  {\mathcal F}(={\mathcal B}_{\overline{\mathbb R}_+}:\mbox{the Borel field}), F_k )$
in
$L^\infty (\Omega, \nu )$ such that
\begin{align*}
&
\qquad
[F_k(\Xi )](\omega )=
\begin{cases}
1
\qquad & (\mbox{ if } V_k(\omega) \in \Xi)
\\
%\frac{1}{2} 
%\frac{1}{2} 
0
\qquad & (\mbox{ if } V_k(\omega) \notin \Xi)
\end{cases}
\\
&
(\forall \omega \in \Omega, \forall \Xi \in
{\mathcal F}
={\mathcal B}_{\overline{\mathbb R}_+}
\mbox{
i.e.,
the Bore field in $X(=\overline{\mathbb R}_+)$
%such that
%the closure $\overline{\Xi}$ is compact in 
%$X$
}
)
\end{align*}
Here we identify
$V_k$ with ${\mathsf O}_k$.
% can be identified.
Further, define the observable
${\mathsf O}=(X,  {\mathcal F}, F )$
in
$L^\infty (\Omega ,\nu)$
such that
\begin{align}
F(\Xi)=\frac{1}{2} \Big( F_1(\Xi)+F_2(\Xi) \Big)
\quad
(\forall \Xi \in {\mathcal F})
\label{eq6}
\end{align}
that is,
\begin{align*}
&
%\qquad
[F(\Xi )](\omega )=
\begin{cases}
1
\qquad & (\mbox{ if } V_1(\omega) \in \Xi, \;\; V_2(\omega) \in \Xi)
\\
%\frac{1}{2} 
1/2
\qquad & (\mbox{ if } V_1(\omega) \in \Xi, \;\; V_2(\omega) \notin \Xi)
\\
%\frac{1}{2} 
1/2
\qquad & (\mbox{ if } V_1(\omega) \notin \Xi, \;\; V_2(\omega) \in \Xi)
\\
%\frac{1}{2} 
0\qquad & (\mbox{ if } V_1(\omega) \notin \Xi, \;\; V_2 (\omega) \notin \Xi)
\end{cases}
\\
&
(\forall \omega \in \Omega, \forall \Xi \in
{\mathcal F}
={\mathcal B}_{\overline{\mathbb R}_+}
\mbox{
i.e.,
the Bore field in $X(=\overline{\mathbb R}_+)$
%such that
%the closure $\overline{\Xi}$ is compact in 
%$X$
}
)
\end{align*}
In what follws,
we shall present the three answers to Problem 1
such that
$$
\underset{\mbox{(in $\S$3.1.1)}}{\fbox{Simple answer}}
\xrightarrow[\mbox{(more strict)}]{}
\underset{\mbox{(in $\S$3.1.2)}}{\fbox{Usual answer}}
\xrightarrow[\mbox{(more strict)}]{}
\underset{\mbox{(in $\S$3.1.3)}}{\fbox{Strict answer}}
$$
which are essentially equivalent,
and thus, these are true.

\vskip1.5cm
\subsubsection{The simplest answer to Problem 1}
\it
Fix any $\omega_0 (\in \Omega )$,
which is assumed to be unknown.
\rm
Consider the measurement
${\mathsf M}_{L^\infty(\Omega, \nu  )} ({\mathsf O}=(X, {\mathcal F} , F ),
S_{[\delta_{\omega_0}]})$. 
Then,
Axiom${}^{\mbox{\scriptsize PCMT}}$\;1
says that
\begin{itemize}
\item[(G1)]
the probability that a measured value
$
\left\{\begin{array}{ll}
V_1(\omega_0)
\\
V_2(\omega_0)
\end{array}\right\}
$
obtained by
the measurement
${\mathsf M}_{L^\infty(\Omega, \nu  )} ({\mathsf O}=(X, {\mathcal F} , F ),
S_{[\delta_{\omega_0}]})$
is given by
$
\left\{\begin{array}{ll}
1/2
\\
1/2
\end{array}\right\}
$.
\end{itemize}
Then,
by the
switching to
$
\left\{\begin{array}{ll}
V_2(\omega_0)
\\
%%%2
V_1(\omega_0)
\end{array}\right\}
$,
you gain
$
\left\{\begin{array}{ll}
V_2(\omega_0) -  V_1(\omega_0)
\\
V_1(\omega_0) - V_2(\omega_0)
\end{array}\right\}
$
dollars.
This means that
the expectation
of the switching
gain
is equal to
\begin{align}
(V_2(\omega_0) - V_1(\omega_0))/2
+
( V_1(\omega_0) - V_2(\omega_0))/2
=
0\mbox{, (which is independent of $\omega_0$)}.
\label{eq7}
\end{align}
This implies that
the swapping is even,
i.e., no advantage and no disadvantage.
\par
Since $\omega_0 (\in \Omega )$ is assumed to be unknown,
the measurement
${\mathsf M}_{L^\infty (\Omega, \nu  )} ({\mathsf O}=(X,  {\mathcal F}, F ),
S_{[\delta_{\omega_0}]})$
is also denoted by
${\mathsf M}_{L^\infty (\Omega, \nu  )} ({\mathsf O}=(X,  {\mathcal F}, F ),
S_{[\ast]})$.
Thus, 
when you obtain a measured value $\alpha$
($\in X $)
by
${\mathsf M}_{L^\infty(\Omega, \nu  )} ({\mathsf O}=(X,  {\mathcal F}, F ),
S_{[\ast]})$
(and
you do not have a way for getting to know whether the money included in the other envelope is more or less than $\alpha$ dollars),
you should conclude that
the swapping is even.
%i.e., no advantage and no disadvantage.
% whichever you choose (keeping or switching).
That is, we can not believe in
the proverb:
{\it "The Other Person's Envelope is Always Greener"}.
%%%%%%%%%%%

\subsubsection{The usual answer to Problem 1}
Define the quasi-product observable
${\mathsf O}\overset{q}{\times} {\mathsf O}$
$=(X\times X, {\mathcal F}\boxtimes {\mathcal F}, F\overset{q}{\times}F )$
in $L^\infty (\Omega, \nu )$
such that
\begin{align*}
&
%\qquad
[
(F {\overset{q}{\times}} F)
(\Xi \times \Gamma )](\omega )=
\begin{cases}
1
\qquad & 
(\mbox{ if } V_1(\omega) \in \Xi, \;\; V_2(\omega) \in \Xi,\;\;  V_1(\omega) \in \Gamma, \;\; V_2(\omega) \in \Gamma )
\\
%\frac{1}{2} 
1/2
\qquad & 
(\mbox{ if } V_1(\omega) \in \Xi, \;\; V_2(\omega) \notin \Xi,\;\;  V_1(\omega) \notin \Gamma, \;\; V_2(\omega) \in \Gamma )
%(\mbox{ if } V_1(\omega) \in \Xi, \;\; V_2(\omega) \notin \Xi)
\\
%\frac{1}{2} 
1/2
\qquad & 
(\mbox{ if } V_1(\omega) \notin \Xi, \;\; V_2(\omega) \in \Xi,\;\;  V_1(\omega) \in \Gamma, \;\; V_2(\omega) \notin \Gamma )
%(\mbox{ if } V_1(\omega) \notin \Xi, \;\; V_2(\omega) \in \Xi)
\\
%\frac{1}{2} 
0\qquad &
(\mbox{ if } V_1(\omega) \notin \Xi, \;\; V_2(\omega) \notin \Xi,\;\;  V_1(\omega) \notin \Gamma, \;\; V_2(\omega) \notin \Gamma )
% (\mbox{ if } V_1(\omega) \notin \Xi, \;\; V_2 (\omega) \notin \Xi)
\end{cases}
\\
&
\qquad
\qquad
(\forall \omega \in \Omega,\;\; \;\;  \forall \Xi, \forall \Gamma \in
{\mathcal F}
={\mathcal B}_{\overline{\mathbb R}_+}
\mbox{
i.e.,
the Bore field in $X(=\overline{\mathbb R}_+)$
%such that
%the closure $\overline{\Xi}$ is compact in 
%$X$
}
)
\end{align*}

%Vxxxxxxxxxxxx
\par
\noindent
\it
Fix any $\omega_0 (\in \Omega )$,
which is assumed to be unknown.
\rm
Consider the measurement
${\mathsf M}_{L^\infty(\Omega, \nu  )} ($
${\mathsf O}\overset{q}{\times} {\mathsf O}$
$=(X\times X, {\mathcal F}\boxtimes {\mathcal F}, F\overset{q}{\times}F ),$
$
S_{[\delta_{\omega_0}]})$. 
Then,
Axiom${}^{\mbox{\scriptsize PCMT}}$\;1
says that
\begin{itemize}
\item[(G2)]
the probability that a measured value
$
\left\{\begin{array}{ll}
(V_1(\omega_0), V_2(\omega_0 ) )
\\
(V_2(\omega_0), V_1(\omega_0))
\end{array}\right\}
$
obtained by
the measurement
${\mathsf M}_{L^\infty(\Omega, \nu  )} ($
${\mathsf O}\overset{q}{\times} {\mathsf O}$
$=(X\times X, {\mathcal F}\boxtimes {\mathcal F}, F\overset{q}{\times}F ),$
$
S_{[\delta_{\omega_0}]})$
is given by
$
\left\{\begin{array}{ll}
1/2
\\
1/2
\end{array}\right\}
$.
\end{itemize}
Here,
\begin{itemize}
\item[(G3)]
"a measured value
$
\left\{\begin{array}{ll}
(V_1(\omega_0), V_2(\omega_0 ) )
\\
(V_2(\omega_0), V_1(\omega_0))
\end{array}\right\}
$
is obtained"
means
\\
"you and the host respectively get
$
\left\{\begin{array}{ll}
V_1(\omega_0)
\\
V_2(\omega_0)
\end{array}\right\}
$
dollars
and
$
\left\{\begin{array}{ll}
V_2(\omega_0)
\\
V_1(\omega_0)
\end{array}\right\}
$dollars"
\end{itemize}
%Then,
%\begin{align*}
%&
%\mbox{
%by the
%switching to
%$
%\left\{\begin{array}{ll}
%(V_2(\omega_0), V_1(\omega_0 ) )
%\\
%(V_1(\omega_0), V_2(\omega_0))
%\end{array}\right\}
%$
%from
%$
%\left\{\begin{array}{ll}
%(V_1(\omega_0), V_2(\omega_0 ) )
%\\
%(V_2(\omega_0), V_1(\omega_0))
%\end{array}\right\}
%$},
%\\
%&
%\mbox{
%you gain
%$
%\left\{\begin{array}{ll}
%V_2(\omega_0) -  V_1(\omega_0)
%\\
%V_1(\omega_0) - V_2(\omega_0)
%\end{array}\right\}
%$
%dollars.}
%\end{align*}
Therefore,
\begin{itemize}
\item[(G4)]
your expectation
$[\frac{V_1(\omega_0)+ V_2(\omega_0)}{2}]$
and
the host's expectation
$[\frac{V_2(\omega_0)+ V_1(\omega_0)}{2}]$
are equal.
%
%
% get
%$
%\left\{\begin{array}{ll}
%V_2(\omega_0) -  V_1(\omega_0)
%\\
%V_1(\omega_0) - V_2(\omega_0)
%\end{array}\right\}
%$
%dollars
%with probability
%$
%\left\{\begin{array}{ll}
%1/2
%\\
%1/2
%\end{array}\right\}
%$
\end{itemize}
%
%
%This means that
%the expectation
%of the switching
%gain
%is equal to
%\begin{align}
%(V_2(\omega_0) - V_1(\omega_0))/2
%+
%( V_1(\omega_0) - V_2(\omega_0))/2
%=
%0\mbox{, (which is independent of $\omega_0$)}.
%\label{eq8}
%\end{align}
This implies that
the swapping is even in the sense of (G4),
i.e., no advantage and no disadvantage.
%Since $\omega_0 (\in \Omega )$ is assumed to be unknown,
%the measurement
%${\mathsf M}_{L^\infty (\Omega, \nu  )} ({\mathsf O}=(X,  {\mathcal F}, F ),
%S_{[\delta_{\omega_0}]})$
%is also denoted by
%${\mathsf M}_{L^\infty (\Omega, \nu  )} ({\mathsf O}=(X,  {\mathcal F}, F ),
%S_{[\ast]})$.
%Thus, 
%when you obtain a measured value $\alpha$
%($\in X $)
%by
%${\mathsf M}_{L^\infty(\Omega, \nu  )} ({\mathsf O}=(X,  {\mathcal F}, F ),
%S_{[\ast]})$
%(and
%you do not have a way for getting to know whether the money included in the other envelope is more or less than $\alpha$ dollars),
%you should conclude that
%the swapping is even.
%i.e., no advantage and no disadvantage.
% whichever you choose (keeping or switching).
That is, we can not believe in
the proverb:
{\it "The Other Person's Envelope is Always Greener"}.
%%%%%%%%%%%2^Xx(\omega_n)xxxxxxxxxxx\Omega_0VVVVVVV2V2V2V_n\omega_n\omega_m
\par
\vskip1.3cm

\subsubsection{The strict answer ((F3): only one measurement is prmitted)}

%Vxxxxxxxxxxxx
\it
\par \noindent
\par
Fix any $\omega_0 (\in \Omega )$,
which is assumed to be unknown.
\rm
Consider the measurement
${\mathsf M}_{L^\infty(\Omega, \nu  )} ($
${\mathsf O}\overset{q}{\times} {\mathsf O}$
$=(X\times X, {\mathcal F}\boxtimes {\mathcal F}, F\overset{q}{\times}F ),$
$
S_{[\delta_{\omega_0}]})$. 
And further, consider
the infinite parallel measurement
$\bigotimes_{n \in {\mathbb N }} {\mathsf M}_{L^\infty(\Omega, \nu  )} ($
${\mathsf O}\overset{q}{\times} {\mathsf O}$
$=(X\times X, {\mathcal F}\boxtimes {\mathcal F}, F\overset{q}{\times}F ),$
$S_{[\delta_{\omega_0}]})$,
whichi is,
by (\ref{eq4}), characterized as follows.
$$
{\mathsf M}_{L^\infty(\Omega^{\mathbb N}, \otimes_{n \in {\mathbb N}
} \nu )} \bigg(
\bigotimes_{n \in {\mathbb N }}({\mathsf O}\overset{q}{\times} {\mathsf O})
=\Big ((X\times X)^{\mathbb N}, \boxtimes_{n \in {\mathbb N}}({\mathcal F}\boxtimes {\mathcal F}),\otimes_{n \in {\mathbb N }}( F\overset{q}{\times}F) \Big),
S_{[\delta_{(\omega_0)_{n \in {\mathbb N}} }]}\bigg)
$$

\par
\noindent
Then,
Axiom${}^{\mbox{\scriptsize PCMT}}$\;1
says that
\begin{itemize}
\item[(G5)]
the probability $P({\widehat \Xi})$ that a measured value
%$
%\left\{\begin{array}{ll}
%(V_1(\omega_0), V_2(\omega_0 ) )
%\\
%(V_2(\omega_0), V_1(\omega_0))
%\end{array}\right\}
%$
obtained by
the infinite parallel measurement
$\bigotimes_{n \in {\mathbb N }} {\mathsf M}_{L^\infty(\Omega, \nu  )} ($
${\mathsf O}\overset{q}{\times} {\mathsf O}$
$=(X\times X, {\mathcal F}\boxtimes {\mathcal F}, F\overset{q}{\times}F ),$
$S_{[\delta_{\omega_0}]})$
belongs to
${\widehat \Xi}
(\in
({\mathcal F}\boxtimes {\mathcal F})^{\boxtimes{\mathbb N}}
)
$
is given by
$$
P({\widehat \Xi})
=
\Big[
\Big(
\bigotimes_{n \in {\mathbb N }}( F\overset{q}{\times}F)
\Big)
({\widehat \Xi})
\Big](\omega_0, \omega_0, \cdots )
$$
\end{itemize}
Here,
put
%\begin{itemize}
%\item[(G1)]
$$
{\widehat \Xi}
=
\Big\{
(x_n, y_n)_{n \in {\mathbb N }}
\in
(X \times X)^{\mathbb N}
\;|
\;
\lim \frac{1}{N} \sum_{n=1}^N
x_n
=
\lim \frac{1}{N} \sum_{n=1}^N
y_n
=
\frac{V_1(\omega_0)+V_2(\omega_0)}{2}
\Big\}
$$
Then we see, by (G2)
and
the law of large numbers ({\it cf.} \cite{Ishi5, Ishi10, Kolm}),  that
\begin{align*}
&
P({\widehat \Xi})
=
\Big[
\Big(
\bigotimes_{n \in {\mathbb N }}( F\overset{q}{\times}F)
\Big)
({\widehat \Xi})
\Big](\omega_0, \omega_0, \cdots )
=1
\\
&
\lim \frac{1}{N} \sum_{n=1}^N
x_n
=
\lim \frac{1}{N} \sum_{n=1}^N
y_n
=
\frac{V_1(\omega_0)+V_2(\omega_0)}{2}
\quad
(\forall 
(x_n, y_n)_{n \in {\mathbb N }}
\in
{\widehat \Xi}
)
\end{align*}
This implies that
the swapping is even,
in the above sense,
% of (\ref{eq8}),
i.e., no advantage and no disadvantage.
%Since $\omega_0 (\in \Omega )$ is assumed to be unknown,
%the measurement
%${\mathsf M}_{L^\infty (\Omega, \nu  )} ({\mathsf O}=(X,  {\mathcal F}, F ),
%S_{[\delta_{\omega_0}]})$
%is also denoted by
%${\mathsf M}_{L^\infty (\Omega, \nu  )} ({\mathsf O}=(X,  {\mathcal F}, F ),
%S_{[\ast]})$.
%Thus, 
%when you obtain a measured value $\alpha$
%($\in X $)
%by
%${\mathsf M}_{L^\infty(\Omega, \nu  )} ({\mathsf O}=(X,  {\mathcal F}, F ),
%S_{[\ast]})$
%(and
%you do not have a way for getting to know whether the money included in the other envelope is more or less than $\alpha$ dollars),
%you should conclude that
%the swapping is even.
%%i.e., no advantage and no disadvantage.
%% whichever you choose (keeping or switching).
That is, we can not believe in
the proverb:
{\it "The Other Person's Envelope is Always Greener"}.
\par
\noindent
\par
It should be noted that
the above explanation is common
to
both
Probalem 1 and Problem 1$'$.
That is, we assert that
$$
\underset{\mbox{(in $\S$1.1)}}{\fbox{Problem 1}}
=
\underset{\mbox{(in $\S$3.1.1)}}{\fbox{Simple answer}}
=
%\xrightarrow[\mbox{(more strict)}]{}
\underset{\mbox{(in $\S$3.1.2)}}{\fbox{Usual answer}}
=
%\xrightarrow[\mbox{(more strict)}]{}
\underset{\mbox{(in $\S$3.1.3)}}{\fbox{Strict answer}}
=
\underset{\mbox{(in $\S$1.1)}}{\fbox{Problem 1$'$}}
$$
Therefore,
quantum language says that
\begin{itemize}
\item[(G6)]
\textcolor{blue}{Problem 1 is essentially the same as Problem 1$'$.}
% are equivalent
\end{itemize}
Also, we believe that
this (G6) is just the excellent statisticians' assertion.
%this (G) is the execellent statisticians' assertion.

%%%%%%%%%%%2^Xx(\omega_n)xxxxxxxxxxx\Omega_0VVVVVVV2V2V2V_n\omega_n\omega_m
\par
%\vskip0.3cm
\vskip0.5cm
\noindent
\bf
Remark 4.
\rm
If
it holds that
$V_1(\omega_0)=V_2(\omega_0)$,
it is clear that
the swapping is even.
Therefore,
without loss of generality,
we can assume that
$V_1(\omega_0) \not=V_2(\omega_0)$.
Also,
it should be noted that
the above argument is applicable to
the simplest case that
$\Omega =\{\omega_0\}$,
i.e.,
the one-point space.
\subsection{The answer in which it is easy to notice the mistake (\ref{eq1})}
\par
\noindent
\par
The answer in Section 3.1 is best, but it does not explain why
we make
% Why do we commit 
a mistake (\ref{eq1}).
% is committed. 
Thus we add the following.
\rm
\subsubsection{The simplest answer to Problem 1}
\par
\noindent
\par
Put
$\Omega=
\{(\omega, 2\omega, ) \;|\;
\omega \in \overline{\mathbb R}_+
\}$.
Here note that
the
$\Omega$ can be identified with
$\overline{\mathbb R}_+$,
i.e.,
\begin{align}
\Omega \ni (\omega, 2\omega ) \longleftrightarrow \omega \in  \overline{\mathbb R}_+
\label{eq8}
\end{align}
and assume that
it has the Lebesgue measure $\nu(d \omega )$,
which is simply denoted by $d \omega$ from here.

Define the observable 
${\mathsf O}=(X(=\overline{\mathbb R}_+),  {\mathcal F}(={\mathcal B}_{\overline{\mathbb R}_+}:\mbox{ the Borel field}), F )$
in
$L^\infty (\Omega, d \omega )$ such that
\begin{align}
&
[F(\Xi )](\omega, 2 \omega )
\Big(\equiv [F(\Xi )](\omega )
\Big)
=
\begin{cases}
1
\qquad & (\mbox{ if } \omega \in \Xi, \;\; 2 \omega \in \Xi)
\\
%\frac{1}{2} 
1/2
%\qquad & (\mbox{ if } V(\omega_n) \in \Xi, \;\; 2V(\omega_n) \notin \Xi)
\qquad & (\mbox{ if } \omega \in \Xi, \;\; 2 \omega \notin \Xi)
\\
%\frac{1}{2} 
1/2
%\qquad & (\mbox{ if } V(\omega_n) \notin \Xi, \;\; 2V(\omega_n) \in \Xi)
\qquad & (\mbox{ if } \omega \notin \Xi, \;\; 2 \omega \in \Xi)
\\
%\frac{1}{2} 
0
%\qquad & (\mbox{ if } V(\omega_n) \notin \Xi, \;\; 2V (\omega_n) \notin \Xi)
\qquad & (\mbox{ if } \omega \notin \Xi, \;\; 2 \omega \notin \Xi)
\end{cases}
\qquad
(\forall (\omega, 2 \omega ) \in \Omega, \forall \Xi \in
{\mathcal F}
%\mbox{
%i.e.,
%$\Xi$
%is
%a Borel subset of $X(=\overline{\mathbb R}_+)$
%such that
%the closure $\overline{\Xi}$ is compact in 
%$X$
%}
)
\label{eq9}
\end{align}
%V
Thus, for any unknown state $(\omega_0, 2 \omega_0 ) (\in \Omega )$,
we have the measurement
${\mathsf M}_{L^\infty (\Omega, d \omega )} ({\mathsf O}=(X,  {\mathcal F}, F ),
$
$
S_{[ \delta_{(\omega_0, 2 \omega_0 )}]})$. 
And, we see,
by Axiom${}^{\mbox{\scriptsize PCMT}}$\;1, that
\begin{itemize}
\item[(H1)]
the probability that a measured value
$x (\in X(=\overline{\mathbb R}_+))$
%=
%\left\{\begin{array}{ll}
%2^m
%\\
%2^{m+1}
%\end{array}\right\}
%$
obtained by
${\mathsf M}_{L^\infty (\Omega, d \omega )} ({\mathsf O}=(X,  {\mathcal F}, F ),$
$
S_{[\delta_{(\omega_0, 2 \omega_0 )}]})$
is equal to
$
\left\{\begin{array}{ll}
\omega_0
\\
2 \omega_0
\end{array}\right\}
$
is given by
$
\left\{\begin{array}{ll}
1/2
\\
1/2
\end{array}\right\}
$.
\end{itemize}
Here,
assume that
$
\left\{\begin{array}{ll}
x =\omega_0
\\
x =2 \omega_0
\end{array}\right\}
$.
Then,
by the
switching to
$
\left\{\begin{array}{ll}
2 \omega_0
\\
\omega_0
\end{array}\right\}
$,
you gain
$
\left\{\begin{array}{ll}
2\omega_0 - \omega_0
\\
\omega_0 - 2\omega_0
\end{array}\right\}
$.
This implies that
the expectation
of the switching
gain
is equal to
$$
%(2^{m+1}-2^m)
(2\omega_0 - \omega_0)
/2
+
%(2^{m}-2^{m+1})
(\omega_0 - 2\omega_0)
/2
=
0
$$
which implies that
the swapping is even,
i.e., no advantage and no disadvantage.
\par
\noindent
\par
Since $(\omega_0, 2 \omega_0)  (\in \Omega )$ is assumed to be unknown,
the measurement
${\mathsf M}_{L^\infty (\Omega, d \omega  )} ({\mathsf O}=(X,  {\mathcal F}, F ),$
$
S_{[\delta_{(\omega_0, 2 \omega_0)}]})$
is also denoted by
${\mathsf M}_{L^\infty (\Omega, d \omega  )} ({\mathsf O}=(X,  {\mathcal F}, F ),
S_{[\ast]})$.
Thus, 
when you obtain a measured value $\alpha$
($\in X $)
by
${\mathsf M}_{L^\infty(\Omega, d \omega  )} ({\mathsf O}=(X,  {\mathcal F}, F ),
S_{[\ast]})$,
you should conclude that
the swapping is even.

\rm
\subsubsection{Why do we make a mistake (\ref{eq1})?}
\par
\unitlength=0.8mm
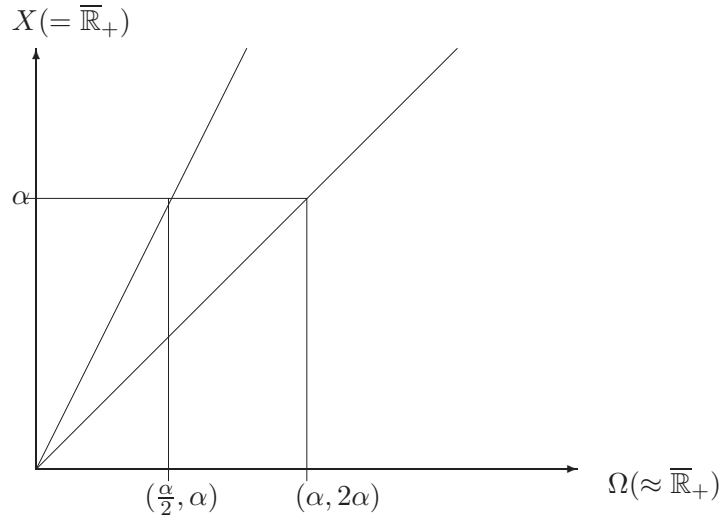
\begin{picture}(140,90)
\put(50,0){{
\put(10,10){\vector(0,1){70}}
\put(10,10){\vector(1,0){90}}
\put(8,55){\path(0,0)(47,0)(47,-47)}
\put(8,55){\path(24,0)(24,-47)}
\put(6,54){$\alpha$}
\put(28,4){$(\frac{\alpha}{2}, \alpha)$}
\put(53,4){$(\alpha, 2 \alpha)$}
\put(6,83){$X(=\overline{\mathbb R}_+)$}
\put(105,6){$\Omega(\approx \overline{\mathbb R}_+)$}
\put(10,10){\line(1,1){70}}
\put(10,10){\line(1,2){35}}
}}
\end{picture}
\noindent
\par
Now we can explain why
we make
% Why do we commit 
a mistake (\ref{eq1}).
Let
${\mathsf M}_{L^\infty (\Omega, d \omega )} ({\mathsf O}=(X,  {\mathcal F}, F ),$
$
S_{[\ast]})$
be the measurement
considered in Section 3.2.1.
Assume that a measured value $\alpha$
is obtained by
${\mathsf M}_{L^\infty(\Omega, d \omega  )} ({\mathsf O}=(X,  {\mathcal F}, F ),
S_{[\ast]})$.
Here, note that,
the likelihood function (\ref{eq3}) is calculated as follows:
\begin{align*}
%%%\label{3fmlm_cor}
f(\alpha, \omega)
\equiv
\inf_{\omega_1 \in \Omega }
\Big[\lim_{\Xi \to \{ \alpha\}, \Xi \ni \alpha, [{F}(\Xi )](\omega_1) \not= 0} \frac{[{F}(\Xi )](\omega)}{
[{F}(\Xi )](\omega_1)
}
\Big]
=
\begin{cases}
1 \quad & (\omega = (\alpha/2, \alpha ) \mbox{ or }(\alpha, 2 \alpha) )
\\
0 &
\mbox{( elsewhere )}
\end{cases}
%\qquad
%(\forall \omega \in \Omega )
\end{align*}
Therefore,
we can infer, by Theorem 1 (Fisher's maximum likelihood method),
that
\begin{itemize}
\item[(H2)]
the unknown state $[\ast]$ is equal to
$(\alpha/2, \alpha )$ or $(\alpha, 2 \alpha )$,
\\
\mbox{$\Big($
if $[\ast]=(\alpha/2, \alpha )$
[resp.
$[\ast]=(\alpha, 2 \alpha )$
], the switching gain is $(\alpha/2-\alpha)$
[resp. $(2\alpha-\alpha)$]
$\Big)$.
}
\end{itemize}

\rm
However, it is not guaranteed that
%we can not assert that
\begin{itemize}
\item[(H3)]
$
\begin{cases}
\mbox{"the probability that $[\ast]=(\alpha/2, \alpha )$"=1/2},
\\
\mbox{"the probability that $[\ast]=(\alpha, 2 \alpha )$"=1/2},
\\
\mbox{"the probability that $[\ast]$ is elsewhere"=0}
\end{cases}
$
\end{itemize}
That is,
the phrase: "\textcolor{blue}{with probability 1/2}" in
[(P1): Why is it paradoxical?]
is wrong,
and therefore, the expectation of the switching gain
"$
%E_{\mbox{\small other}}
E_{\mbox{\small other}}(\alpha)
-\alpha=(1/2)(\alpha/2) + (1/2)(2\alpha) -\alpha=
\alpha/4 > 0$"
is wrong.
That is, it is impossible to calculate the expected value $E_{\mbox{\small other}}(\alpha)$.
In other words, the expected value $E_{\mbox{\small other}}(\alpha)$ in the formula (\ref{eq1}) is meaningless.

%\vskip0.5cm
\par
\noindent
%44444444444444444
\section{Bayesian approach to the two envelopes paradox }
\par
\noindent
\par
\rm

In the framework of
the pure measurement
${\mathsf M}_{L^\infty(\Omega, d \omega  )} ({\mathsf O}=(X,  {\mathcal F}, F ),
S_{[\ast]})$
(defined in Section 3.2),
we can not derive the statement (H3). 
Thus, next,  consider another situation
of Problem 1 (Bayesian approach to the two envelopes paradox),
i.e.,
the statistical measurement
${\mathsf M}_{L^\infty (\Omega, d \omega )} ({\mathsf O}=(X,  {\mathcal F}, F ),
S_{[\ast]}(\rho_0))$. 
Recalling the identification
(\ref{eq8}):
$\Omega \ni (\omega, 2\omega ) \longleftrightarrow \omega \in  \overline{\mathbb R}_+$,
assume that
$$
\rho_0(D) =\int_D h(\omega ) d \omega 
\quad
(\forall D \in {\mathcal B}_{\Omega }={\mathcal B}_{\overline{\mathbb R}_+ })
$$
where
the probability density function
$h: \Omega ( \approx \overline{\mathbb R}_+ ) \to
\Omega ( = \overline{\mathbb R}_+ )$
is assumed to be continuous positive function.
That is, the mixed state $\rho_0 (\in {\mathcal M}^m(\Omega(=\overline{\mathbb R}_+ ) ) )$ has the probability density function
$h$.

%%Axiom

Axiom$^{\rm \small SCMT}$\;1 says that
\begin{itemize}
\item[(I1)]
The probability $P(\Xi)$
$(\Xi \in {\mathcal B}_X ={\mathcal B}_{\overline{\mathbb R}_+ })$
that
a measured value
%The sample measure $\rho_1$ on $X (=\overline{\mathbb R}_+ )$
obtained by
the statistical measurement
${\mathsf M}_{L^\infty (\Omega, d \omega )} ({\mathsf O}=(X,  {\mathcal F}, F ),
S_{[\ast]}(\rho_0))$
belongs to
$\Xi (\in {\mathcal B}_X ={\mathcal B}_{\overline{\mathbb R}_+ })$
is given by
\begin{align}
P (\Xi )
&
=
\int_\Omega [F(\Xi )](\omega ) \rho_0 (d \omega )
=
\int_\Omega [F(\Xi )](\omega )  h (\omega ) d \omega 
\nonumber
\\
&
=
\int_{\Xi} 
%\chi{{}_{{}_{\Xi}}}
%(\omega )
%\cdot
%\Big(
\frac{h(x/2 )}{4}
+
\frac{h(x )}{2}  
%\Big)
\;\;
d x
\quad
(\forall \Xi \in {\mathcal B_{\overline{\mathbb R}_+ }})
\label{eq10}
\end{align}
\end{itemize}
Therefore, the expectation is given by
\begin{align}
\int_{\overline{\mathbb R}_+} x P(d x )
=
\frac{1}{2}
\int_{0}^\infty
x
\cdot
\Big(
h(x/2 )/2
+
h(x )
\Big)
d x
=
\frac{3}{2}
\int_{\overline{\mathbb R}_+} x h(x ) d x
\label{eq11}
\end{align}
Further,
Theorem 2 (
Bayes' theorem ) says that
\begin{itemize}
\item[(I2)]
When
a measured value
$\alpha$
%The sample measure $\rho_1$ on $X (=\overline{\mathbb R}_+ )$
is obtained by
the statistical measurement
${\mathsf M}_{L^\infty (\Omega, d \omega )} ({\mathsf O}=(X,  {\mathcal F}, F ),$
$
S_{[\ast]}(\rho_0))$,
then
the post-state
$\rho_{\mbox{\scriptsize post}}^\alpha
(\in {\mathcal M}^m (\Omega ))$
is given by
\begin{align}
(\ref{eq5})=
\rho_{\mbox{\scriptsize post}}^\alpha
=
\frac{
\frac{h(\alpha/2)}{2}
}
{\frac{h(\alpha/2)}{2} 
%\delta_{\alpha/2}
+
h(\alpha) 
}
\delta_{(\frac{\alpha}{2}, \alpha)}
+
\frac{
%\frac{f(\alpha/2)}{2} \delta(\omega -\alpha/2)
%+
h(\alpha) 
%\delta(\omega -\alpha)
}
{\frac{h(\alpha/2)}{2} 
%\delta_{\alpha/2}
+
h(\alpha) 
}
\delta_{({\alpha}{},2 \alpha)}
\label{eq12}
\end{align}
\end{itemize}
Hence,

\begin{itemize}
\item[(I3)]
if
$[\ast]
=
$
$
\left\{\begin{array}{ll}
\delta_{(\frac{\alpha}{2}, \alpha)}
\\
\delta_{({\alpha}{}, 2 \alpha)}
\end{array}\right\}
$,
then
we have to change
$
\left\{\begin{array}{ll}
\alpha \longrightarrow \frac{\alpha}{2}
\\
\alpha \longrightarrow 2{\alpha}
\end{array}\right\}
$,
and thus we get the switching gain
$
\left\{\begin{array}{ll}
\frac{\alpha}{2} - \alpha
(= - \frac{\alpha}{2} )
\\
2{\alpha} - \alpha
(= {\alpha})
\end{array}\right\}
$.
\end{itemize}
Therefore, the expectation of the switching gain is calculated as follows:
\begin{align}
&
\int_{\overline{\mathbb R}_+}  
\Big(
(-\frac{\alpha}{2})
\frac{
\frac{h(\alpha/2)}{2}
}
{\frac{h(\alpha/2)}{2} 
%\delta_{\alpha/2}
+
h(\alpha) 
}
%\delta_{(\frac{\alpha}{2}, \alpha)}
+
\alpha
\frac{
%\frac{f(\alpha/2)}{2} \delta(\omega -\alpha/2)
%+
h(\alpha) 
%\delta(\omega -\alpha)
}
{\frac{h(\alpha/2)}{2} 
%\delta_{\alpha/2}
+
h(\alpha) 
}
\Big)
P(d \alpha )
\nonumber
\\
=
&
\int_{\overline{\mathbb R}_+} 
%\int_{\Xi} 
%\chi{{}_{{}_{\Xi}}}
%(\omega )
%\cdot
%\Big(
(-\frac{\alpha}{2})\frac{h(\alpha/2 )}{4}
+
\alpha \cdot \frac{h(\alpha )}{2}  
%\Big)
\;\;
d \alpha
=0
%\frac{3}{2}
%\int_{\overline{\mathbb R}_+} \omega h(\omega ) d \omega
%%\delta_{({\alpha}{},2 \alpha)}
\label{eq13}
\end{align}
Therefore,
if
$\int \omega h(\omega ) d \omega < \infty $,
we see, by (\ref{eq11}),
that
the swapping is even,
i.e., no advantage and no disadvantage,
in the sense of (\ref{eq13}).

%555555555555555555555555

\section{The St. Petersburg two-envelope paradox in quantum language}

%\section{The quantum linguistic description of the St. 
%Petersburg two-envelope paradoxes}

\subsection{The St. Petersburg two-envelope paradox}
\par
\noindent
\par
In what follows, we introduce the St. Petersburg two-envelope problem ({\it cf.} {\cite{Chal}}),
which is well known
as a kind of high school students' mathematical puzzle.
%Let us rewrite Problem 2 as follows.

\par
\noindent
{\bf Problem 2}
\rm
[The St. Petersburg two envelope problem].
You are presented with two envelopes, A and B. You are told that each of them contains an amount determined by the following procedure, performed separately for each envelope: a coin was flipped until it came up heads, and if it came up heads on the nth trial, $2^n$ is put into the envelope. This procedure is performed separately for each envelope. You are given envelope A, and you find 
$2^m$ dollars in the envelope $A$.
Now you are offered the options of keeping A or switching to B. What should you do? 
\par \vskip0.3cm
\par
\noindent
[(P2);Why is it paradoxical?].
You reason that, before opening the envelopes A and B, the expected values
$E(x)$ and $E(y)$ in A and B is infinite respectively.
For any $2^m$, if you knew that A contained $x=2^m$ dollars, then the expected value $E(y)$ in B would still be infinite. 
%So (2) for all $2^m$, if you knew that A contained $2^m$, you 
%would have an expected gain in switching to B. 
Therefore,
you should switch to B. But this seems clearly wrong, \textcolor{blue}{as your information about A and B is symmetrical}. 
This is the famous St. Petersburg two-envelope paradox
(i.e.,
"The Other Person's Envelope is Always Greener"
).

\par
\noindent
\subsection{(P2): The St. Petersburg two-envelope paradox
in Statistical CMT ( without Bayes' method)}
%The quantum linguistic description in non-Bayesian statistics}
\par
\noindent
\par
Here,
let us explain the St. Petersburg two-envelope paradox
in Statistical CMT  ( without Bayes' method).

\par
\noindent
\par
Define the state space
$\Omega$ such that
$\Omega=\{\omega \; |\; \omega=1,2, \cdots \}$
with the counting measure
$\nu$,
that is, 
the set of all natural numbers.
And define the observable
${\mathsf O}=(X,  {\mathcal F}, F )$
such that
\begin{align*}
&
X= \{ k \;|\; k =1,2 , \cdots  \},
\quad {\mathcal F}=2^{X}
%\;|\; \mbox{$\Xi$ is a finite set} \}
% \quad {\mathcal F}=2^{X}
\\
&
[F(\Xi)](\omega)
= 
\begin{cases}
1 \quad & (\mbox{ if } \omega \in \Xi )
\\
0 \quad & (\mbox{ elsewhere })
\end{cases}
\qquad
(\forall \Xi \in {\mathcal F}, \forall \omega =1,2, \cdots )
\end{align*}
%Fix any $m (\in {\mathbb Z} )$,
%which is assumed to be unknown.
Define the mixed state $\rho_0$
(i.e., the probability measure on $\Omega$)
such that
$$
\rho_0 (\{\omega \})=
\begin{cases}
1/2^m  \quad & (\mbox{ if } \omega = 2^m, m=1,2,... )
\\
0 \quad & (\mbox{ elsewhere })
\end{cases}
$$
Consider the statistical measurement
${\mathsf M}_{L^\infty(\Omega, \nu )} ({\mathsf O}=(X,  {\mathcal F}, F ),
S_{[\ast]}(\rho_0) )$. 
Axiom${}^{\mbox{\scriptsize SCMT}}$\;1 
says that
\begin{itemize}
\item[(J1)]
the probability that a measured value
$x (\in X)$
obtained by
${\mathsf M}_{L^\infty (\Omega, \nu )} ({\mathsf O}=(X,  {\mathcal F}, F ),
S_{[\ast]}(\rho_0))$
is equal to
$
2^k
$
is given by
$
2^{-k}
$.
\end{itemize}
Therefore, the expectation $E(x)$ of a measured value
is equal to
$$
E(x)= \sum_{k=1}^\infty 2^k \cdot 2^{-k}= \infty
$$
Now consider the parallel measurement
${\mathsf M}_{L^\infty (\Omega \times \Omega , \nu \otimes \nu)} ({\mathsf O} \otimes {\mathsf O}=(X \times X,  {\mathcal F} \boxtimes  {\mathcal F} , F \otimes F ),
S_{[(\ast, \ast)]}(\rho_0 \otimes \rho_0 ))$,
where
$\rho_0 \otimes \rho_0 $ is the product measure on $\Omega \times \Omega$.
%Recalling Definition 1,
By the similar way of Definition 1,
we consider that
this parallel measurement
is the same as
taking a measurement
${\mathsf M}_{L^\infty (\Omega, \nu )} ({\mathsf O}=(X,  {\mathcal F}, F ),
S_{[\ast]}(\rho_0))$
twice.
Let
$(x,y)(\in X \times X )$
be the measured value
obtained by
the parallel measurement
${\mathsf M}_{L^\infty (\Omega \times \Omega, \nu \otimes \nu )} ({\mathsf O} \otimes {\mathsf O}=(X \times X,  {\mathcal F} \boxtimes  {\mathcal F} , F \otimes F ),
S_{[\delta_{(\omega_0, \omega_0)}]})$.
We of course see that
the expectation $E(x,y)=(\infty, \infty )$.
Problem 2 says that
you got a measured value $2^m$
(i.e.,
$2^m$ dollars in Envelope A).
Namely,
$x=2^m$.
However,
since $E(y)=\infty$,
in the next measurement,
you are expected to get a measured value $y$
such that
$E(y)=\infty  > 2^m$.
That is, you are expected to find $y$ dollars (i.e.,
$E(y)=\infty  > 2^m$)
in Envelope B.
Thus, you should switch to Envelope B.
\par
\vskip0.5cm
\par
\noindent
\bf
Remark 5.
\rm
(i):
\rm
Note that,
in the above argument,
Axiom$^{\mbox{\scriptsize SCMT}}$\;1 
is used, and not Bayes' theorem (Theorem 2)
 ({\it cf.} the formula (\ref{eq5})).
\par
\noindent
(ii):
Recall the statement
"\textcolor{blue}{as your information about A and B is symmetrical}"
in [(P2): Why is it paradoxical?].
This statement is not true, since you find $2^m$ dollars in Envelope A,
but
you do not open Envelope B yet.
Therefore, Problem 2 is not paradoxical.
That is, we can believe in
the proverb:
{\it "The Other Person's Envelope is Always Greener"}.
However,
if you do not open both envelopes,
Envelopes A and B are even.
\par
\noindent
(iii): 
%In the above sense, we have to reconsider even the statement
%"\textcolor{blue}{as your information about A and B is symmetrical}"
%in [(P1): Why is it paradoxical?].
%True reason should be explained such as mentioned in
%[(P1): The quantum linguistic description].
%\par
%\noindent
%(iii):
The probability $P(y>2^m)$ such that
"$y >2^m$"
is easily calculated as follows.
$$
P(y>2^m)= \frac{1}{2^m}
$$
Concerning the St. Petersburg two-envelope paradox,
this "probability criterion" may be rather reasonable.

\par
\noindent
\subsection{(P2): The St. Petersburg two-envelope paradox
in Pure CMT ($\approx$ non-Bayesian statistics)}
%The quantum linguistic description in non-Bayesian statistics}

\par
\noindent
\par
Let us explain the St. Petersburg two-envelope paradox
in Pure CMT,
which is essentially the same as
the argument in the previous section 5.2.

\par
\noindent
\par
Define the state space
$\Omega$ such that
$\Omega=\{\omega_0\}$,
that is, 
the set composed of one element,
where
$nu(\{\omega_0\})=1$.
And define the observable
${\mathsf O}=(X,  {\mathcal F}, F )$
such that
\begin{align*}
&
X= \{ 2^k \;|\; k =1,2 , \cdots  \}, 
\quad {\mathcal F}=2^{X}\;
%|\; \mbox{$\Xi$ is a finite set} \}
%\quad {\mathcal F}=2^{X}
\\
&
[F(\{2^k\})](\omega_0)
= 2^{-k}
\qquad
(k=1,2, \cdots )
\end{align*}
%Fix any $m (\in {\mathbb Z} )$,
%which is assumed to be unknown.
Consider the measurement
${\mathsf M}_{L^\infty (\Omega, \nu )} ({\mathsf O}=(X,  {\mathcal F}, F ),
S_{[\delta_{\omega_0}]})$. 
Then, Axiom${}^{\mbox{\scriptsize PCMT}}$\;1 
says that
\begin{itemize}
\item[(J2)]
the probability that a measured value
$x (\in X)$
obtained by
${\mathsf M}_{L^\infty (\Omega, \nu )} ({\mathsf O}=(X,  {\mathcal F}, F ),
S_{[\delta_{\omega_0}]})$
is equal to
$
2^k
$
is given by
$
2^{-k}
$.
\end{itemize}
Therefore, the expectation $E(x)$ of a measured value
$x$
is equal to
$$
E(x)= \sum_{k=1}^\infty 2^k \cdot 2^{-k}= \infty
$$
Now consider the parallel measurement
${\mathsf M}_{L^\infty (\Omega \times \Omega, \nu \otimes \nu )} ({\mathsf O} \otimes {\mathsf O}=(X \times X,  {\mathcal F} \boxtimes  {\mathcal F} , F \otimes F ),
S_{[\delta_{(\omega_0, \omega_0)}]})$.
Recalling Definition 1,
we consider that
this parallel measurement
is the same as
taking a measurement
${\mathsf M}_{L^\infty (\Omega, \nu )} ({\mathsf O}=(X,  {\mathcal F}, F ),
S_{[\delta_{\omega_0}]})$
twice.
Let
$(x,y)(\in X \times X )$
be the measured value
obtained by
the parallel measurement
${\mathsf M}_{L^\infty (\Omega \times \Omega, \nu \otimes \nu )} ({\mathsf O} \otimes {\mathsf O}=(X \times X,  {\mathcal F} \boxtimes  {\mathcal F} , F \otimes F ),
S_{[\delta_{(\omega_0, \omega_0)}]})$.
We of course see that
the expectation $E(x,y)=(\infty, \infty )$.
Problem 2 says that
you got a measured value $2^m$
(i.e.,
$2^m$ dollars in Envelope A).
Namely,
$x=2^m$.
However,
since $E(y)=\infty$,
in the next measurement,
you are expected to get a measured value $y$
such that
$E(y)=\infty  > 2^m$.
That is, you are expected to find $y$ dollars (i.e.,
$E(y)=\infty  > 2^m$)
in Envelope B.
Thus, you should switch to Envelope B.
\par
\vskip0.5cm
\par
\noindent
\bf
Remark 6.
\rm
The above answer may be rather fit for the following problem.
\begin{itemize}
\item[(K)]
In the envelope A and the envelope B,
there are infinite pins $\{ P_k \}_{k=1}^\infty$ with the length $2^{-k}$.
The pin $P_k$ may be identified with the interval
$(2^{-k}, 2^{1-k} ]$
$(\subseteq  (0,1])$.
%(2^{1-k}-2^{-k} (= ))$
%$(k=1,2, \cdots)$.
Assume that the pin $P_k$ is $2^k$ dollars.
And further assume that the probability that a pin $P_k$ will be picked from Envelope A (and Envelope B) is given by $2^{-k}$.
You are given envelope A, and, from the envelope $A$, you pick up a pin $P_m$, which is 
$2^m$ dollars.
Now you are offered the options of keeping A or switching to B. What should you do? 
\end{itemize}
Although this and Problem 2 are similar but somewhat different,
we consider that two answers (in Sections 5.2 and 5.3) are valid.

%%6666666666666666

\par

%\vskip0.5cm
\par
\par
\noindent
\section{Conclusions}
\par

\noindent
\par
In order to show the great descriptive power of quantum language
(i.e.,
"quantum language is future statistics"),
we want to assert that
\begin{itemize}
\item[(L1)]
if a probabilistic problem is
described in terms of quantum language,
the problem will be automatically solved.
\end{itemize}
As one of examples (L1),
in this paper
we showed that
the two envelope problem is automatically solved
in Section 3 (non-Bayesian two envelopes paradox)
and
Section 4 (Bayesian two envelopes paradox).

\par
The readers may ask the following question:
\begin{itemize}
\item[(L2)]
Why is it hard to make a mistake in quantum language method?
\end{itemize}
We consider that this is due to the fact:
\begin{itemize}
\item[(L3)]
Quantum language has visible key-words:
"measurement",
"observable",
"state",
"measured value".
And these concepts are motivated by quantum mechanics.
\\
On the other hand,
statistics has invisible key-words:
"probability space",
"random variable",
"parameter".
\end{itemize}
This is our answer to the question (L2).
Also, it should be noted that
the sum (\ref{eq6}) of observables
has not appeared once throughout our research
\cite{Ishi3}-\cite{Ishi100},
that is, it is rare in the usual situations.
In this sense, the two envelopes problem may be
tricky and paradoxical.
After all,
we conclude that
quantum language provides the final answer
(i.e.,
the easiest and deepest understanding 
)
to the two envelope-problem.

Also, we add:
\begin{itemize}
\item[(L4)]
In Section 5, we see that
the St. Petersburg two-envelope paradox has two
formulations
(i.e.,
Classical SMT and Classical PMT),
and also,
the St. Petersburg two-envelope paradox is independent of Bayes' method,
and thus it is not related to Bayesian statistics ({\it cf.} Remark 2).
\end{itemize}
For completeness,
our main assertion (G6) is again rewritten as follows.
\begin{itemize}
\item[(L5)]
quantum language says that,
if Problem 1 is a scientific statement,
Problem 1 should be essentially the same as Problem 1$'$.
If the reader wants to assert that these are different,
he has to propose another language (except quantum) by which
Problem 1 and Problem 1$'$
are described as the different problems.
That is because
we believe Wittgenstein's words
(i.e.,
the spirit of the philosophy of language):"The limits of my language mean the limits of my world."
\end{itemize}
\par
We hope that our proposal will be discussed and examined from various view-points.

\rm
%\vskip-0.5cm
\par
%\vskip0.2cm 

%\end{document}
%EEEEEEEEEEEEEEEEEEEEEEEEEE

%%\
%6. References}
%\section{References}
{
\small

\normalsize
}

\end{document}